\documentclass[journal ]{new-aiaa}
\usepackage[utf8]{inputenc}
\usepackage{textcomp}

\usepackage{graphicx}
\usepackage{float}   
\usepackage{amsmath}
\usepackage[version=4]{mhchem}
\usepackage{siunitx}
\usepackage{longtable,tabularx}
\usepackage{adjustbox}
\usepackage{booktabs}

\title{Transonic buffet and incompressible low-frequency oscillations at high Reynolds numbers}

\author{Vishw Patel, Aman Jain and Jewel Rupini\footnote{Masters students, Department of Aerospace Engineering.}}
\affil{Indian Institute of Technology Kanpur, Kanpur, Uttar Pradesh, 208016, India}
\author{Antony Raja Arulsekar\footnote{Undergraduate student, Department of Mechanical Engineering.}}
\affil{National Institute of Technology, Tiruchirappalli, Tamil Nadu, 620015, India}
\author{Pradeep Moise\footnote{Assistant Professor, Department of Aerospace Engineering, pmoise@iitk.ac.in, pradeep890@gmail.com (Corresponding author).}}
\affil{Indian Institute of Technology Kanpur, Kanpur, Uttar Pradesh, 208016, India}

\begin{document}

\maketitle

\begin{abstract}
Coherent, self-sustained oscillations of the flow over aircraft wings can lead to unsteady loads that detrimentally affect aircraft safety and stability, thus limiting the flight envelope. Two such types of oscillations are the low-frequency oscillations (LFO) observed in flow over airfoils close to stall in the incompressible regime and transonic buffet, which occurs at high speeds and involves oscillating shock waves. The possibility that these two are linked has been explored only recently at low Reynolds numbers ($\boldsymbol{Re\sim O}\mathbf{(10^4)}$) and natural transition conditions  (Moise \textit{et al.}, \textit{J. Fluid Mech.}, vol. 981, 2024, p. A23). However, the shock wave structure in the transonic regime under these conditions differs substantially when compared to high Reynolds number flows, and it is unknown whether a connection can be established at high Reynolds numbers. This study investigates this possibility by performing incompressible and compressible URANS simulations at $\boldsymbol{Re} = \mathbf{10^7}$. We show that transonic buffet exists for a narrow range of freestream Mach numbers across a wide range of angles of attack ($\boldsymbol{\alpha}$), and that buffet-like oscillations are observed at higher $\boldsymbol{\alpha}$ even in the absence of shock waves. Using a spectral proper orthogonal decomposition (SPOD), we show that the dominant modes associated with these oscillations are strongly correlated for all cases, even in the absence of shock waves. Furthermore, using a fully incompressible URANS framework, we capture LFO at the same Reynolds number and confirm the connection between these two phenomena using SPOD. These results imply that neither shock waves nor compressibility is necessary to sustain such low-frequency oscillations, suggesting that the fundamental mechanism governing them is related to flow separation. This can potentially help in improved control strategies to extend the flight envelope by mitigating buffet or LFO.
\end{abstract}

\section*{Nomenclature}

\noindent(Nomenclature entries should have the units identified)

{\renewcommand\arraystretch{1.0}
\noindent\begin{longtable*}{@{}l @{\quad=\quad} l@{}}
$C_p$& time-averaged pressure coefficient \\
$C_f$& time-averaged skin friction coefficient \\
$C_L$& instantaneous lift coefficient from URANS \\
$\bar{C}_L$& time-averaged lift coefficient \\
$C_L'$& fluctuating component of lift coefficient \\
$\Delta C_L$ & Difference between maximum and minimum $C_L$ past transients (${C_L}_\mathrm{max}-{C_L}_\mathrm{min}$)\\
$x$& chordwise direction \\
$y$& wall-normal direction \\
$\Delta y^+$& grid spacing adjacent to the wall in wall units \\
$c$   & chord \\
$t$ & dimensionless time \\
$t_0$ & dimensionless time taken for transient evolution \\
d$t$ & dimensionless time step \\
$T_b$ & dimensionless buffet time period \\
$T_\mathrm{up}$ & time taken by upstream propagating waves starting from the trailing edge to reach the shock \\
$T_\mathrm{down}$ & time taken by downstream propagating waves starting from the shock foot to reach the trailing edge \\
$\rho$ & density\\
$\alpha$ & angle of attack\\
$M$ & freestream Mach number\\
$M_\mathrm{loc}$ & local Mach number\\
$a$ & speed of sound \\
$Re$ & freestream Reynolds number\\
$\Omega$ & spatial domain\\
$\boldsymbol{x}$, $\boldsymbol{x^{\prime}}$ & points in the spatial domain\\
$\boldsymbol{W}$ & weights associated with grid points\\
$St$ & Strouhal number\\
$St_b$ & Strouhal number associated with buffet-like oscillations\\
$\boldsymbol{S}$& cross-spectral density tensor \\ 
$\lambda$ & eigenvalue of dominant mode from spectral proper orthogonal decomposition (SPOD)\\
$\boldsymbol{\psi}(\boldsymbol{x},St)$ & SPOD mode at a specific frequency, $St$\\
$\boldsymbol{\phi}(\boldsymbol{x},t)$& spatio-temporal SPOD mode \\
$\boldsymbol{W}$& weight associated with inner product in SPOD (based on cell volume) \\
$\chi$ & phase of SPOD mode, $\boldsymbol{\psi}$, at a given frequency \\
$n$ & number of grid points\\
$p$ & pressure field \\
$k$ & turbulent kinetic energy \\
$\omega$ & specific dissipation rate of $k$\\


\end{longtable*}}

\section{Introduction\label{secIntro}}
\lettrine{V}{arious} types of self-sustained, coherent, flow oscillations that can cause load fluctuations and structural vibrations have been reported when aircraft fly at different flow conditions \cite{Hilton1947,Zaman1989, Dandois2018, Zauner2020FTaC}. At low speeds and in the incompressible regime,  low-frequency oscillation (LFO) has been observed for high angles of attack just below the stall angle, characterized by regular boundary layer separation and reattachment \cite{Zaman1989, Sandham2008, Almutairi2010, Mons2024}. Similarly, in the high-speed transonic regime, transonic buffet has been identified at low angles of attack (including $\alpha = 0^\circ$), characterized by large-scale fore-aft shock wave motion \cite{McDevitt1985,Lee1989,Zauner2019}. Such oscillations are detrimental to flight performance as they can lead to structural damage and loss of control, thus limiting the flight envelope and maneuverability. For these reasons, understanding the mechanisms and mitigating incompressible LFO and transonic buffet has been the focus of intense research. However, other than the recent study of Moise \textit{et al.} \cite{Moise2024} at low freestream Reynolds numbers, most studies have investigated these two phenomena in isolation and have not fully explored the possibility that they could be connected. In the present study, we examine the relation between the two by performing numerical simulations of the flow over an airfoil in different regimes at high Reynolds numbers. 

Zaman \textit{et al.} \cite{Zaman1989} were among the first to experimentally study the phenomenon of LFO in low-speed flows over airfoils at high angles of attack close to stall. They observed that LFO is distinct from "bluff-body shedding" (\textit{i.e.}, von Kármán vortex street, which occurs at a relatively higher frequency that is one or two orders higher than that of LFO and dominates at angles of attack above stall), and that LFO is characterized by "a periodic switching between stalled and unstalled states". At high Reynolds numbers, LFO occurs close to stall and coexists with vortex shedding for the same inflow conditions due to hysteresis effects \cite{Hristov2018, Atallah2024}. Even in the fully post-stall regime, LFO can be excited by active flow control techniques \cite{Bernardini2016}. Various simulation approaches have captured LFO, including those employing panel methods coupled with integral boundary-layer equations \cite{Sandham2008}, unsteady Reynolds-averaged Navier-Stokes (URANS) simulations \cite{Iorio2016}, and three-dimensional direct numerical simulations \cite{Almutairi2010, Moise2024}. 

In high-speed flows, transonic buffet on airfoils is usually identified by large-scale shock wave motion in the fore-aft direction \cite{Lee1990}. However, like LFO, quasi-periodic flow separation and reattachment are also characteristic features of transonic buffet. Different types of transonic buffet have been observed based on flow conditions. For zero and low angles of attack, a Type I buffet occurs, characterized by shock wave motion on both sides of the airfoil \cite{Iovnovich2012, Giannelis2017, Moise2022}. At higher angles of attack, the flow oscillations are dominant only on the suction side, referred to as Type II buffet. Independently, buffet can be classified based on the Reynolds number and the type of boundary layer transition achieved. At high Reynolds numbers ($Re \sim O(10^6)$ or higher, for $Re$ based on chord and freestream conditions), or when the boundary layer transition is forced in the vicinity of the leading edge, a single oscillating shock wave terminates the supersonic region, referred to as turbulent transonic buffet. By contrast, at lower Reynolds numbers and natural transition conditions, multiple shock waves occur and oscillate \cite{Zauner2019}. Although the amplitude of shock wave oscillation can be highly sensitive to transition conditions \cite{Lusher2024}, the spatio-temporal structure of the oscillatory modes appears to be qualitatively similar irrespective of the shock wave structure \cite{Moise2022Trip}. Furthermore, a different instability, associated with the laminar separation-bubble, which causes load fluctuations, can also accompany buffet for natural transition conditions \cite{Brion2020, Dandois2018, Zauner2023, Zauner2024}. It should also be emphasized that the dominant unsteady dynamics can be significantly different when a sweep angle is imposed  \cite{Crouch2019, Paladini2019a} or when considering flow over three-dimensional swept wings \cite{Dandois2016, Timme2020} due to the prevalence of three-dimensional instabilities (buffet cells). However, the focus of this study is only on two-dimensional buffet on airfoils. This phenomenon remains relevant as it is the dominant instability in several situations. For example, it is observed to be the dominant instability in the experiments \cite{Jacquin2009} on an unswept large-aspect ratio wing. Similarly, the two-dimensional instability is found to be dominant at moderate Reynolds numbers in the large-eddy simulations (LES) of unswept and swept wing configurations of an infinite wing section (aspect ratio of 1 and sweep angles of up to 40$^\circ$) studied by Moise \textit{et al.} \cite{Moise2022} and unswept configurations of Lusher \textit{et al.} \cite{Lusher2025}, although three-dimensional features are also present.

Significant progress has been made in understanding the physical mechanisms underlying LFO and transonic buffet. Among the first to study LFO, Zaman \textit{et al.} \cite{Zaman1989} suggested that it likely originates as a hydrodynamic instability. This has been confirmed using a global linear stability analysis using base flows obtained from the RANS approach \cite{Iorio2016, Busquet2021}. Studies have also shown that the formation of a separation bubble is an important feature for the onset of LFO, and it has been variously related to bubble bursting or shear layer flapping \cite{rinoie_takemura_2004, Aniffa2023a, Aniffa2023b, Eljack_2024}. The origin of transonic buffet has been examined from various standpoints. A popular model for transonic buffet predicts that it is associated with a feedback mechanism that involves waves generated at the shock foot traveling downstream in the boundary layer, impinging on the trailing edge, propagating upstream outside the boundary layer, and interacting with the shock wave \cite{Lee1990}. Experimental and numerical evidence to support or refute this model remains ambiguous, with some studies contradicting it \cite{Jacquin2009, Paladini2019, Moise2022, Moise2022Trip} while others support it \cite{Deck2005, Xiao2006a, Mahji2023}. By contrast, Crouch \textit{et al.} \cite{Crouch2007, Crouch2009, Crouch2024} used a global linear stability analysis of RANS base states, with the eddy viscosity terms also linearized, to show that transonic buffet is associated with a supercritical Hopf bifurcation of a globally unstable mode. While this gives important insights into the origin of transonic buffet, exploring aspects such as the cause of this instability and the roles of the shock wave and boundary layer separation can lead to a better physical understanding of the phenomenon. In this regard, while many early studies assumed that shock waves are essential for these instabilities to occur, recent studies such as that of Paladini \textit{et al.} \cite{Paladini2019} suggest that the role of shock waves is only secondary to transonic buffet based on a sensitivity analysis.


By performing Large-eddy simulations (LES) of transonic buffet at low Reynolds numbers ($Re = 50,000$), zero angle of attack, and natural transition conditions (\textit{i.e.}, laminar buffet), Moise \textit{et al.} \cite{Moise2024} were able to show that while Type I transonic buffet occurs at high freestream Mach numbers of $M \geq 0.8$ with shock waves present in the flow field, similar buffet-like oscillations persist at lower freestream Mach numbers of $M = 0.72$ even though the flow remains entirely subsonic. This further corroborates the conclusion of Paladini \textit{et al.} \cite{Paladini2019a} that shock waves are secondary to transonic buffet. Furthermore, starting from transonic buffet conditions with oscillating shock waves and performing LES at progressively higher angles of attack and lower freestream Mach numbers, Moise \textit{et al.} were able to show that the oscillations are sustained at high angles of attack even when the local Mach number was substantially low ($M_\mathrm{loc} \leq 0.6$) and most regions of the flow could be approximated as incompressible, indicating that the oscillations observed are associated with the LFO phenomenon. This was the first study to establish a connection between transonic buffet and LFO. However, the study focused on low Reynolds numbers and free transition conditions, where the shock wave structure is significantly different from that observed at high Reynolds numbers. Thus, the authors concluded that studies at high Reynolds numbers are required to generalize their observations. Additionally, the study did not perform simulations at fully incompressible conditions to link transonic buffet with LFO.

Motivated by these gaps in the connections between transonic buffet and incompressible LFO, the present study investigates the relation between the two phenomena at high Reynolds numbers by performing simulations using the URANS framework (both compressible and incompressible). The rest of the study is organized as follows. The numerical methodology adopted is explained in Sec.~\ref{secMethods}, while the results of the compressible and incompressible RANS simulations are provided in Sec.~\ref{secCompressibleRANS} and Sec.~\ref{secIncompressibleRANS}, respectively. The results of the simulations are analyzed using a spectral proper orthogonal decomposition (SPOD) in Sec.~\ref{secSPOD}. The implications of the results are discussed in Sec.~\ref{secDiscussion} and the conclusion is provided in Sec.~\ref{secConclusion}. 

\section{Methodology\label{secMethods}}

\subsection{Governing equations and flow setup \label{subSecFlowSetup}}
In the present study, the two-dimensional compressible and incompressible RANS equations are numerically solved to simulate the flow of a Newtonian fluid over the symmetric NACA 0012 airfoil. The working fluid is assumed to be air, and in the compressible simulations, it is treated as calorically perfect, with its dynamic viscosity satisfying Sutherland's law, while in the incompressible simulations, density and viscosity are constant. Compressible URANS simulations were performed to simulate transonic buffet, which occurs at high freestream Mach numbers at low angles of attack. Following this, fresh compressible simulations were run with the freestream Mach number reduced and the angle of attack increased such that the oscillations are sustained (i.e., by estimating and staying within the range of buffet onset and offset Mach numbers for each angle of attack studied). By successively increasing the angle of attack while reducing the freestream Mach numbers, we were able to sustain oscillations at low freestream Mach numbers that are close to the incompressible limit. In addition, simulations were also performed using the framework of fully incompressible RANS equations to capture LFO. The choice of the turbulence model for the RANS equations is discussed in detail in Sec.~\ref{subSecMethodTurbModels}. 

The NACA 0012 airfoil with a blunt trailing edge of thickness 0.25\% chord is studied here. The airfoil is considered to be an adiabatic, no-slip wall. A structured C-grid topology was employed to study the flow over this airfoil. The computational domain for the C-block extends radially to a distance of ten chord lengths. As shown in Sec.~\ref{MethodsValid}, this domain extent is sufficient to capture buffet features, with good agreement between present results and other experiments and simulations. A pressure outlet boundary condition is applied at the outer boundary of the C block, while the freestream flow variables, such as velocity and temperature, are set to achieve the desired freestream Mach and Reynolds numbers. Most cases were studied by performing steady RANS simulations for the conditions associated with that case, where we start by using a standard initialization based on the freestream conditions. The initial conditions for the URANS simulations are based on the saturated solutions obtained from these steady RANS simulations. It was observed that the steady RANS simulations do not converge at high angles of attack for the compressible simulations when the SST $k-\omega$ model is used. For these cases alone, the URANS simulations were directly initialized without performing steady RANS simulations. Unless explicitly mentioned, the initial conditions of the results reported in the subsequent sections are based on the steady RANS approach.

\subsection{Turbulence models}
\label{subSecMethodTurbModels}
Transonic buffet is well-known to be sensitive to the turbulence model used in URANS simulations (see Sec.~3.1.1 in \cite{Giannelis2017}), although the main drawback seems to be the inaccurate prediction of the onset conditions. Additionally, when there is boundary layer separation and large separated regions (which occur in both LFO and transonic buffet), the current state-of-the-art turbulence models for closure of RANS equations can be inadequate, and high-fidelity approaches such as detached-eddy or large-eddy simulations are preferred for better accuracy. Nevertheless, the focus of this study is not on accurately predicting buffet onset or the exact conditions required for buffet. Rather, it is on examining its spatio-temporal features under whichever conditions it occurs. Thus, we adopted the numerically less-expensive RANS approach, with the turbulence models chosen as the standard Menter's Shear-stress transport (SST) k-$\omega$ and Spalart-Allmaras (SA) models \cite{Menter, Spalart}. Both have been shown to be effective for simulating transonic buffet \cite{Giannelis2017}, while the latter is commonly used to simulate incompressible LFO (e.g., \cite{Iorio2016, Busquet2021}). It has been suggested that the coefficient, $a_1$, in Menter's SST $k-\omega$ model needs to be reduced to achieve unsteadiness at experimentally-reported onset conditions for ONERA's OAT15A airfoil at moderate Reynolds numbers \cite{Giannelis2018}. However, at the high Reynolds numbers studied here for the NACA 0012 airfoil, buffet features were found to be insensitive to $a_1$, with the results for the unmodified standard value adequately matching experimental observations (see Sec.~\ref{MethodsValid}). Thus, we have used the standard values for all parameters of this turbulence model \cite{Menter}. 

As noted before, the compressible simulations were first performed at low angles of attack at high freestream Mach numbers, where transonic buffet is observed, followed by fresh simulations at successively higher angles of attack at lower freestream Mach numbers, towards incompressible LFO. For transonic buffet occurring at low angles of attack and high freestream Mach numbers, no significant differences were observed between results obtained using SST $k-\omega$ and SA models and previous experimental studies \cite{McDevitt1985}, as shown in Sec.~\ref{MethodsValid}. Hence, we opted to use the SST $k-\omega$ model for all compressible simulations, as it is expected to perform better when strong separation develops, which is especially the case for conditions well above buffet onset. However, it was observed that when simulating the compressible or incompressible equations at high angles of attack of $\alpha \geq 18^\circ$ (former, at low freestream Mach numbers), the steady RANS simulation did not converge when the SST $k-\omega$ model was used. When URANS simulations were directly initialized (i.e., without using steady RANS results as initial conditions) and run at these conditions, the regular flow oscillations were found to exhibit an intermittent long-time behavior for conditions just above onset (see Appendix, Fig.~\ref{figApp:LongTime}(a)), while at higher angles of attack, the simulation was found to diverge. By contrast, the use of the SA model leads to regular periodic oscillations at all times and angles (including offset). It is unclear why there is a difference in the behaviors observed for different turbulence models. One possible explanation could be that this behavior is due to the experimentally reported bistability of LFO with vortex shedding in the post-stall regime, which can lead to intermittent spontaneous switching between flow states \citep{Hristov2018, Atallah2024}. Another possibility could be numerical instabilities. Nevertheless, for both SST $k-\omega$ and SA models, coherent oscillations at frequencies relevant to buffet and LFO are observed, as discussed in detail in Sec.~\ref{secIncompressibleRANS} and Sec.~\ref{secSPOD}. Hence, unless otherwise specified, the results from compressible simulations are based on the SST $k-\omega$ model. For incompressible simulations, similar features such as intermittent oscillations and numerical divergence persist for the SST $k-\omega$ model (see Appendix, Fig.~\ref{figApp:LongTime}(b)), and results based on both models are compared and discussed, and the model used is explicitly specified wherever incompressible results are described.

\subsection{Numerical setup\label{subSecNumSetup}}

\begin{figure}[hbt!]
    \centering
    \includegraphics[trim={0cm 0.5cm 0cm 1.5cm}, clip,width=0.495\textwidth]{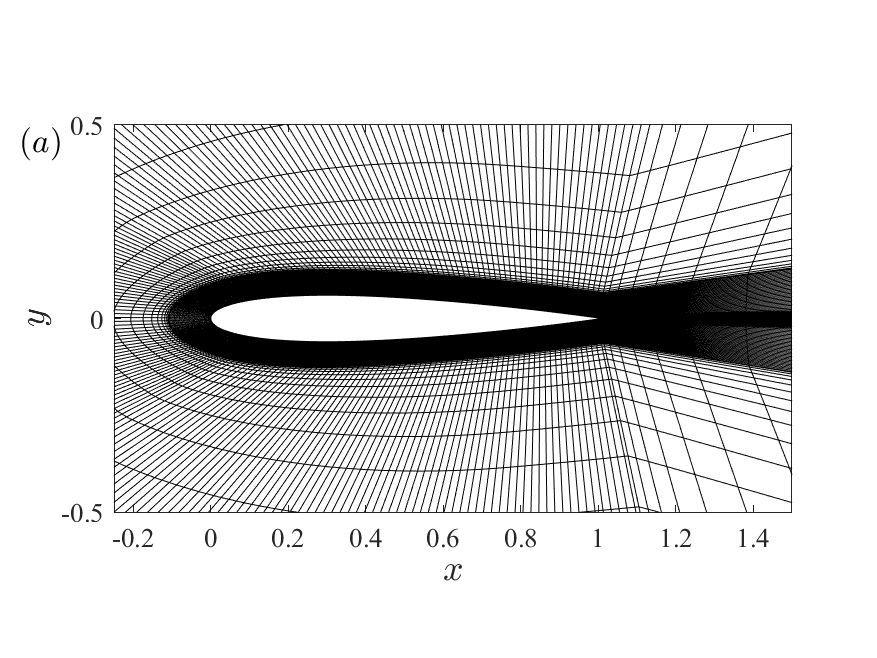}
    \includegraphics[trim={0cm 0.5cm 0cm 1.5cm}, clip,width=0.495\textwidth]{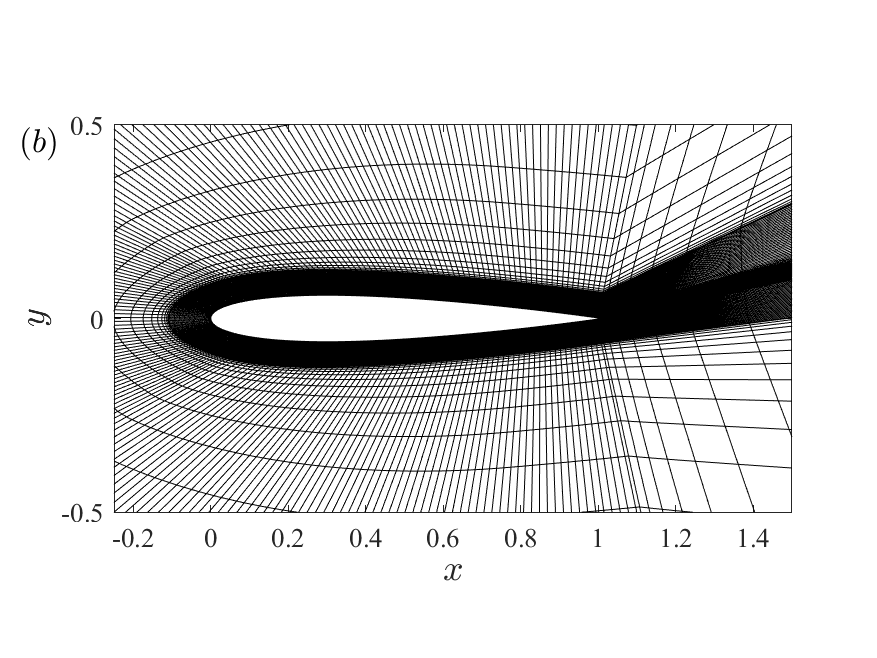}
    \caption{Grids used for flow simulations at (a) low ($\alpha \leq 9^\circ$), and (b) high angles of attack. For clarity, only every second grid point is shown. }
    \label{fig:grid}
\end{figure}

The simulations are carried out using the commercial flow solver, Ansys Fluent 2021 R1, with the RANS equations solved using a cell-centered, finite-volume method. The implicit density-based solver is used to perform the compressible flow simulations, and Roe's flux difference splitting scheme is used to evaluate inviscid fluxes \cite{Roe}. By contrast, the pressure-based solver is chosen for incompressible flow simulations. In both cases, the second-order upwind scheme and the least-squares cell-based approach are used for interpolation at the cell faces and to evaluate gradients, respectively.  A second-order, implicit dual-time stepping formulation is used for temporal discretization with a fixed dimensionless physical time step, scaled using freestream velocity and chord, of $\approx 0.05$ for all compressible cases (approximately 200 iterations per buffet cycle) and $0.1$ for incompressible simulations (approximately 150 iterations per LFO cycle). 

 The variations in angle of attack for the different cases are implemented by modifying the freestream velocity components as required, while the grid remains the same. However, since the orientation of the wake is dependent on the angle of attack, two different grids are used for low and high angles of attack, such that the wake region is resolved. Fig.~\ref{fig:grid}(a) and Fig.~\ref{fig:grid}(b) show the symmetric grid used at low and high angles of attack, respectively. The former is used for angles of attack from 4$^\circ$ to 9$^\circ$, while the latter is used for angles from 10$^\circ$ to 21$^\circ$. 
These curvilinear, structured, multi-block grids consist of 93682 nodes with a node distribution of 460 $\times$ 200 along the airfoil surface and in the wall-normal direction, respectively. For comparison, Crouch \textit{et al.} \cite{Crouch2009} used a distribution of 455 $\times$ 145 for their RANS simulations at the same flow conditions. In the present work, the wall-normal spacing in wall units was confirmed \textit{a posteriori} to be less than unity, \textit{i.e.}, $\Delta y^+ < 1$.

\subsection{Parametric variation\label{subSecParams}}
In the present study, the parameters varied include $M$ (based on freestream velocity and temperature scales) and the angle of attack of the NACA 0012 airfoil at a fixed Reynolds number of $Re = 10^7$ (based on freestream velocity, density and viscosity scales, and the airfoil chord). Compressible simulations (using Menter's SST $k-\omega$ model) were performed by varying $\alpha$ in steps of $1^\circ$ in the range $4^\circ \leq \alpha \leq 19^\circ$, and finding at least one $M$ at which unsteady lift oscillations at a low frequency are observed. A summary of these cases, which are the focus of discussions in the subsequent sections, is provided in Table.~\ref{tab:cases}. Additionally, to determine the onset and offset conditions of buffet with $M$ at a fixed $\alpha$, multiple simulations were carried out at different $M$. For $4^\circ \leq \alpha \leq 8^\circ$, this was examined in steps of $1^\circ$, but at higher angles of attack in the range $10^\circ \leq \alpha \leq 16^\circ$, these simulations were carried out only in steps of $2^\circ$ to reduce numerical expense. For $\alpha > 16^\circ$, it was not possible to determine the offset boundary due to intermittent oscillations and divergence issues, as detailed in Sec.~\ref{secCompressibleRANS}.

For the incompressible simulations, $\alpha$ is varied to determine the onset and offset of LFO. Both SA and SST $k-\omega$ models were tested. For the former case, simulations were carried out at $\alpha = 19^\circ$ and by varying $\alpha$ in steps of $0.25^\circ$ in the range of $20^\circ \leq \alpha < 22^\circ$. Oscillatory solutions were observed only for $20.5^\circ \leq \alpha \leq 21.5^\circ$, and the case of $\alpha = 21^\circ$ is discussed here in detail. For the SST $k-\omega$ model, simulations were performed in the same range as SA, and LFO was observed at $\alpha = 22^\circ$, but intermittent oscillations and divergence issues prevented examination of higher angles of attack, and the offset angle of attack could not be determined.

\begin{table}[H]
\caption{\label{tab:cases} Flow parameters and results for different angles of attack studied.}
\centering
\begin{tabular}{ccccc|ccccc}
\toprule
\hline
 $\alpha$ & M & $\bar{C}_{L}$ & $St_b$ & $\Delta C_L$ & 
  $\alpha$ & M & $\bar{C}_{L}$ & $St_b$ & $\Delta C_L$ \\ 
\hline
$4^\circ$ & 0.76 & 0.459 & 0.057 & 0.399 & 
$13^\circ$ & 0.39 & 1.080 & 0.117 & 0.662 \\
$5^\circ$ & 0.71 & 0.615 & 0.067 & 0.224 & 
$14^\circ$ & 0.37 & 1.134 & 0.123 & 1.011 \\
$6^\circ$ & 0.68 & 0.703 & 0.075 & 0.311 & 
$15^\circ$ & 0.35 & 0.964 & 0.095 & 1.39 \\
$7^\circ$ & 0.65 & 0.730 & 0.078 & 0.551 & 
$16^\circ$ & 0.33 & 1.109 & 0.102 & 1.827 \\
$8^\circ$ & 0.60 & 0.737 & 0.089 & 0.644 & 
$17^\circ$ & 0.30 & 1.28 & 0.051 & 1.921 \\
$9^\circ$ & 0.59 & 1.036 & 0.092 & 0.571 & 
$18^\circ$ & 0.28 & 1.41 & 0.148 & 1.554 \\
$10^\circ$ & 0.50 & 0.904 & 0.095 & 0.415 & 
$19^\circ$ & 0.26 & 1.44 & 0.159 & 1.257 \\
$11^\circ$ & 0.47 & 0.926 & 0.105 & 0.564 &
$21^\circ$ (SA) & Inc. & 1.460 & 0.052 & 1.694\\
$12^\circ$ & 0.42 & 1.030 & 0.113 & 0.959  &
$22^\circ$ (SST) & Inc. & 0.919 & 0.12 & 0.907\\
\hline
\bottomrule
\end{tabular}
\end{table}

\subsection{Spectral proper orthogonal decomposition\label{MethodsSPOD}}

Spectral proper orthogonal decomposition is used to extract spatio-temporally coherent features from the flow-field data \cite{Lumley1970, Glauser1987, Towne2018}. This approach was employed by Moise \textit{et al.} \cite{Moise2024} to connect transonic buffet and LFO at low Reynolds numbers and is also adopted here for consistency and comparison purposes. A brief overview of SPOD is provided below (see \cite{Moise2022} for further details). For a stationary, zero-mean, stochastic spatiotemporal process, an optimal basis for representing an ensemble of its realizations is formed by the eigenfunctions $\boldsymbol{\psi}$ of the cross-spectral density tensor, which satisfy
\begin{equation}
\int_{\Omega} \boldsymbol{S}(\boldsymbol{x},\boldsymbol{x^{\prime}},St)\boldsymbol{W}\boldsymbol{\psi}(\boldsymbol{x},St)d\Omega = \lambda(St)\boldsymbol{\psi}(\boldsymbol{x^{\prime}},St).
\end{equation}
 
The temporal evolution of this SPOD mode is given by
\begin{equation}
    \boldsymbol{\phi}(\boldsymbol{x},t) = \operatorname{Re}\{\boldsymbol{\psi}(\boldsymbol{x},St)\exp(2\pi \mathrm{i}\: St \:t)\}.\label{eqnSPODMode}
\end{equation}

The numerical code provided in \cite{SCHMIDT201998} is used for computing the SPOD modes. The cross-spectral density matrix is numerically evaluated using snapshots of simulation data, comprising the two velocity components, pressure, and density obtained from URANS simulations. The snapshots were stored at time intervals of approximately 0.2 (dimensionless, based on chord and freestream velocity), implying a dimensionless sampling frequency of approximately 5. The snapshots were grouped in blocks such that each block contained at least three buffet/LFO cycles. A minimum of 1400 snapshots were used for all cases. Welch's method was used to compute the cross-spectral density matrix, \(\mathbf{S}\), with the Hamming window function and a 50\% overlap between the chosen blocks. The weight for the inner product is chosen as the volume of the cell.

SPOD provides an energy-ranked set of modes for each frequency studied. In the present study, it was observed that the majority of the energy is concentrated in the most energetic mode at all relevant frequencies. Thus, the discussions below are based on this dominant SPOD mode. The pressure field of this SPOD mode at the frequency associated with buffet is used to compare different cases. Note that the SPOD mode at a given frequency, $\boldsymbol{\psi}(\boldsymbol{x},St)$, is a complex field, as described by Eq.~\ref{eqnSPODMode}. The spatio-temporal SPOD mode, $\boldsymbol{\phi}(\boldsymbol{x},t)$, is obtained by finding the real part of $\boldsymbol\psi$ at a given phase, $\chi = St\times t$, of the oscillation cycle. For visual comparisons, a reference phase, $\chi_0$, is required to extract a spatial field from the spatio-temporal mode. Previous studies have found that there is no ideal reference phase and that it is better to choose the phase manually based on visual similarity \cite{Moise2024}. Thus, here we choose an arbitrary phase manually. 
To quantify the relation between any two SPOD modes, we have calculated the magnitude of the Pearson correlation coefficient given by 
\begin{equation}
\text{Pearson correlation coefficient} =  \frac{\left|\sum\limits_{j=1}^n p_{1,j}^{ *} \cdot W_{1,j} \cdot p_{2,j} \cdot W_{2,j}\right|}{\sqrt{\sum\limits_{j=1}^n p_{1,j}^{*} \cdot W_{1,j} \cdot p_{1,j} \cdot W_{1,j}} \sqrt{\sum\limits_{j=1}^n p_{2,j}^{*} \cdot W_{2,j} \cdot p_{2,j} \cdot W_{2,j}}} .
\end{equation}
where \(p_1\) and \(p_2\) are the pressure fields of the two SPOD modes (\textit{i.e.}, eigenvectors) under consideration and $j$ represents the index of the spatial locations. As the SPOD pressure field is complex, the absolute value of the correlation coefficient is considered. Since most of the energy associated with buffet oscillations is concentrated near the airfoil, we have chosen the inner product to be restricted to a region whose extent (scaled using $c$) in the $x-$ and $y-$directions is given by [-1,2] and [-1,1], respectively. 
A correlation value of unity indicates that the two SPOD modes are identical, while a correlation value close to zero implies that the two SPOD modes are uncorrelated and hence, dissimilar.

\subsection{Validation of simulations and grid convergence studies\label{MethodsValid}}

\begin{figure}[t]
    \centering
        \includegraphics[trim={0cm 0cm 0cm 0cm},  
 clip,width=0.45\textwidth]{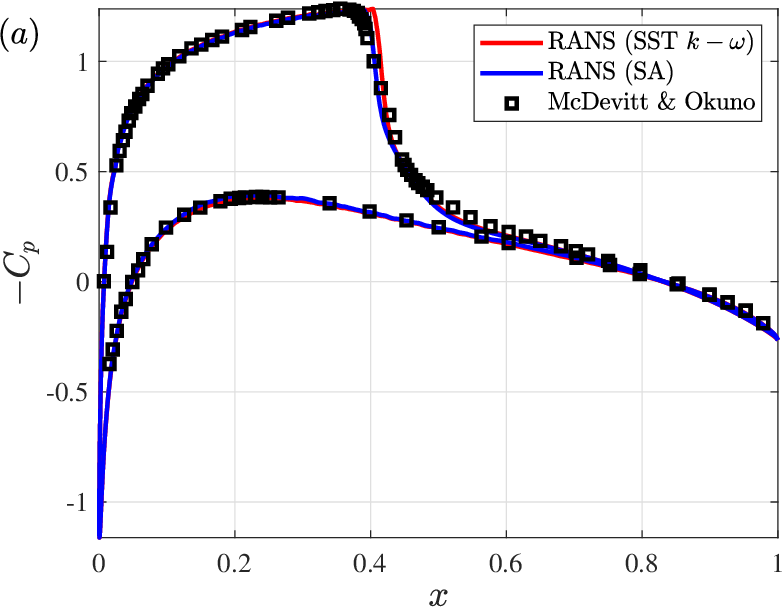}
        \hspace{0.25cm}
        \includegraphics[trim={0cm 0cm 0cm 0cm}, clip,width=0.45\textwidth]{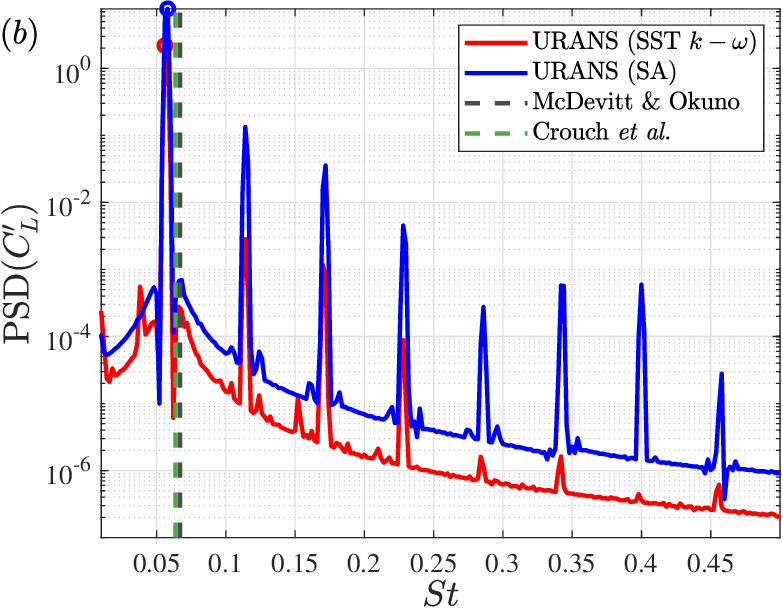}
    \caption{(a) Coefficient of pressure computed in steady RANS simulations at $\alpha = 2^\circ$, $M = 0.75$ and (b) power spectra of the fluctuating lift coefficient in URANS simulations at $\alpha = 4^\circ$, $M = 0.76$. Results from different turbulence models are compared with previously reported results \cite{McDevitt1985,Crouch2009}, for NACA 0012 airfoil at $Re = 10^7$.}
    \label{fig:SAvSST}
\end{figure}

Both steady and unsteady RANS simulations were conducted to investigate the sensitivity of buffet to turbulence models and compare with previously reported experimental \cite{McDevitt1985} and computational results \cite{Crouch2009}. The mean pressure coefficient on the airfoil surface obtained in the steady RANS simulations at conditions of $\alpha = 2^\circ$ and $M = 0.75$ (below buffet onset) with the SA and Menter's SST $k-\omega$ turbulence models is plotted in Fig.~\ref{fig:SAvSST}(a). Note that buffet is not observed under these conditions in experiments as well. It is seen that there is good agreement with experimental results reported at the same flow conditions for both turbulence models, with the mean streamwise location of the shock found to be at $x/c = $ 0.36, 0.39, and 0.40 for the experiments, SA, and SST $k-\omega$ cases, respectively. Spectral characteristics of the lift coefficient at conditions of $\alpha = 4^\circ$ and $M = 0.76$, where buffet is observed, are compared in Fig. \ref{fig:SAvSST}(b). Symbols are used to highlight the peaks in the power spectral density (PSD) associated with transonic buffet and are compared with results reported previously in experiments \cite{McDevitt1985} and URANS simulations using the SA model \cite{Crouch2007} (dashed vertical lines). The buffet Strouhal number for all cases is in good agreement, with the URANS simulations in the present study that employ SST $k-\omega$ and SA model, having peaks at $St_b = 0.056$ and 0.058, respectively, compared to $St_b = 0.063$ reported for URANS simulations with SA model by Crouch \textit{et al.} \cite{Crouch2009}, and $St_b = 0.07$ reported in experiments \cite{McDevitt1976}. Grid convergence was also studied for this case by performing simulations on a refined grid with approximately 1.6 times the number of nodes of the original grid (totaling 153384 nodes), and it was observed that the mean lift coefficient and the Strouhal number were negligibly affected ($|\Delta St| \approx 0.0002$). 

Based on the above results, it can be inferred that the present numerical setup is adequate in capturing the dynamic characteristics of transonic buffet at the flow conditions of $Re = 10^7$. At higher angles of attack (and lower freestream Mach numbers), there is no data available for validation, and thus, experimental studies are required to test the accuracy of the results presented here for such conditions. However, we emphasize that the focus of this study is only on the buffet characteristics and not on accurately predicting its onset or the range of conditions it occurs. The trend observed here is consistent with that at low Reynolds numbers \cite{Moise2024}, as shown in Sec.~\ref{secSPOD}, while SPOD modes are topologically similar to those reported in other studies \cite{Moise2022Trip}, which adds further confidence in the reliability of these results. Similarly, for incompressible LFO, there are no results available for validation or verification under these conditions and only at lower $Re$ (e.g., \cite{Zaman1989,Hristov2018}). We have performed incompressible URANS at lower Reynolds numbers for the NACA 0012 airfoil, where experimental data is available for LFO \cite{Zaman1989} and found the Strouhal number of the oscillations agreeing well with that reported in the experiments (not shown).

\section{Transonic buffet and buffet-like subsonic flow oscillations}
\label{secCompressibleRANS}

\begin{figure}[t!]
    \centering
        \includegraphics[trim={0cm 0cm 0cm 0cm}, clip,width=0.47\textwidth]{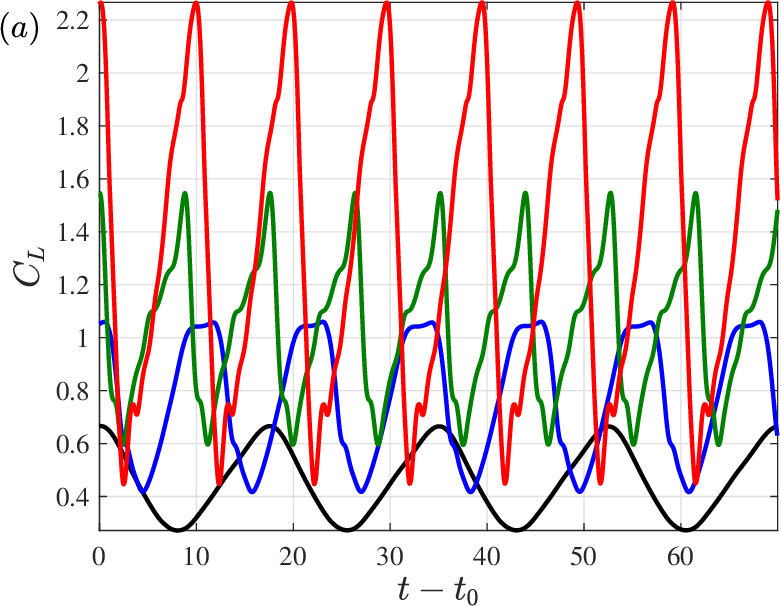}
    \hspace{0.5cm}
        \includegraphics[trim={0cm 0cm 0cm 0cm}, clip,width=0.47\textwidth]{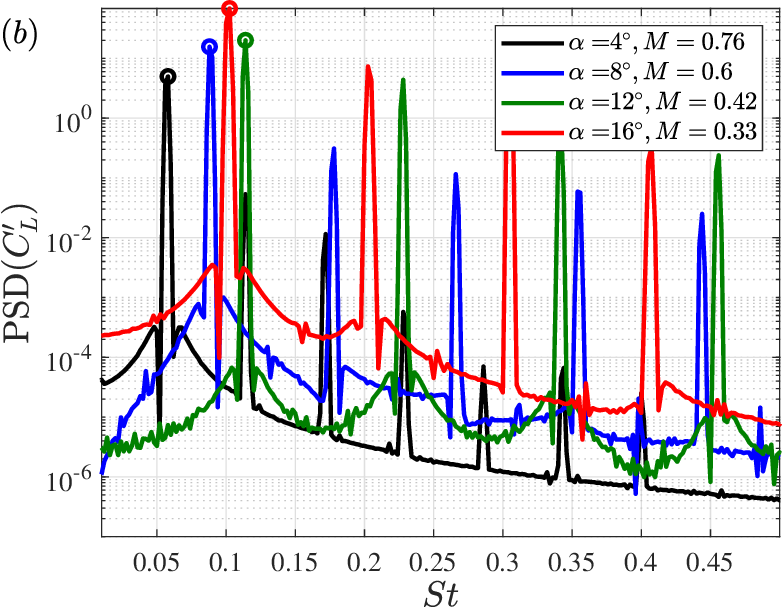}
    \caption{(a) Temporal variation of the lift coefficient past transients and (b) corresponding PSD of its fluctuating component from compressible URANS simulations of the NACA 0012 airfoil at \(Re = 10^7\).}
    \label{fig:Buffet_ClPSD}
\end{figure}

The results obtained from the compressible URANS simulations are presented in this section. The results for only a few select angles of attack are shown for brevity, although simulations were performed at increments of $\Delta \alpha = 1^\circ$ in the range of $4^\circ\leq\alpha\leq19^\circ$ and buffet-like oscillations were observed for all angles in this range (see Table~\ref{tab:cases} and Fig.~\ref{fig:OnOffset}(a)). The freestream Mach number at each $\alpha$ was varied to find a representative case where oscillations at a low frequency are present. The temporal variations of the lift coefficient for these cases are plotted in Fig.~\ref{fig:Buffet_ClPSD}(a) while the corresponding power spectral densities are compared in Fig. \ref{fig:Buffet_ClPSD}(b). The spectral peaks associated with the oscillations at the low frequency are highlighted using circles. The amplitude and frequency associated with these peaks are provided in Table~\ref{tab:cases}. It is evident from the plots that large-amplitude lift oscillations at comparable frequencies ($0.06 < St_b < 0.1$) sustain in both transonic and subsonic regimes. In the transonic regime, these oscillations characterize transonic buffet, as is typically seen in most studies \cite{Lee2001}. The oscillations under other conditions are not considered to be transonic buffet, as no shock wave is present. However, we will collectively refer to oscillations in any regime as `buffet' (or `buffet-like oscillations' when referring specifically to the subsonic regime), as they all exhibit similar characteristics. Note that both $M$ and $\alpha$ are varied simultaneously here, as buffet only occurs in a narrow range of $M$ for a given $\alpha$ (see Fig.~\ref{fig:OnOffset}(a)). Thus, the variations in the amplitude and frequencies across cases do not exhibit any specific trend, as both $M$ and $\alpha$ are expected to strongly influence buffet characteristics \cite{Giannelis2018, Moise2022}. For the entire set of compressible cases studied, the variation in buffet frequency is in the approximate range $0.05 \leq St_b \leq 0.15$ (see Table~\ref{tab:cases}), which is still an order lower than the frequency of modes related to vortex shedding observed on airfoils ($St \approx 1$). 
  
\begin{figure}[H]
    \centering
        \includegraphics[trim={0.2cm 0.15cm 1.25cm 0.7cm}, clip,width=0.425\textwidth]{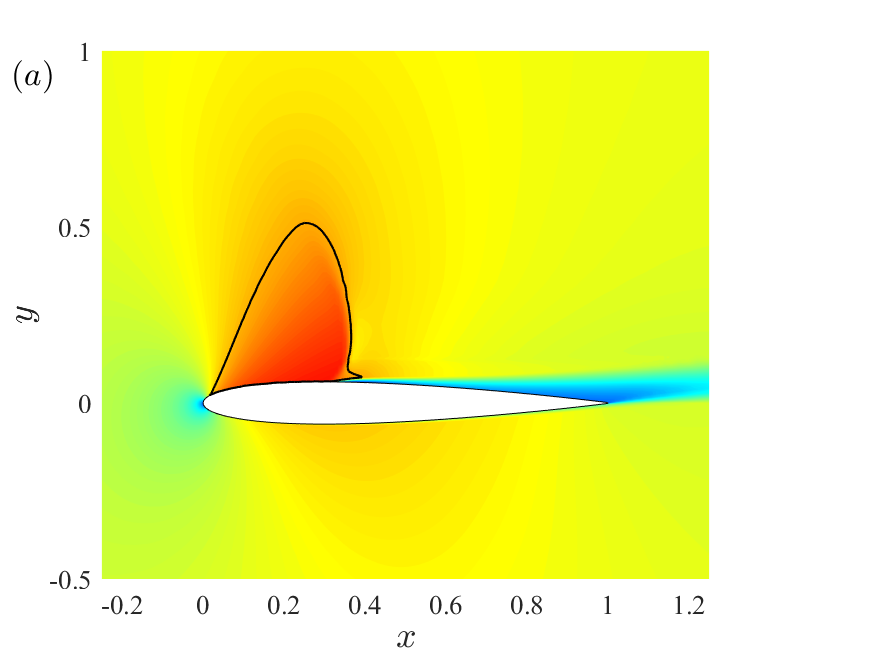}
        \includegraphics[trim={0.2cm 0.15cm 1.25cm 0.7cm}, clip,width=0.425\textwidth]{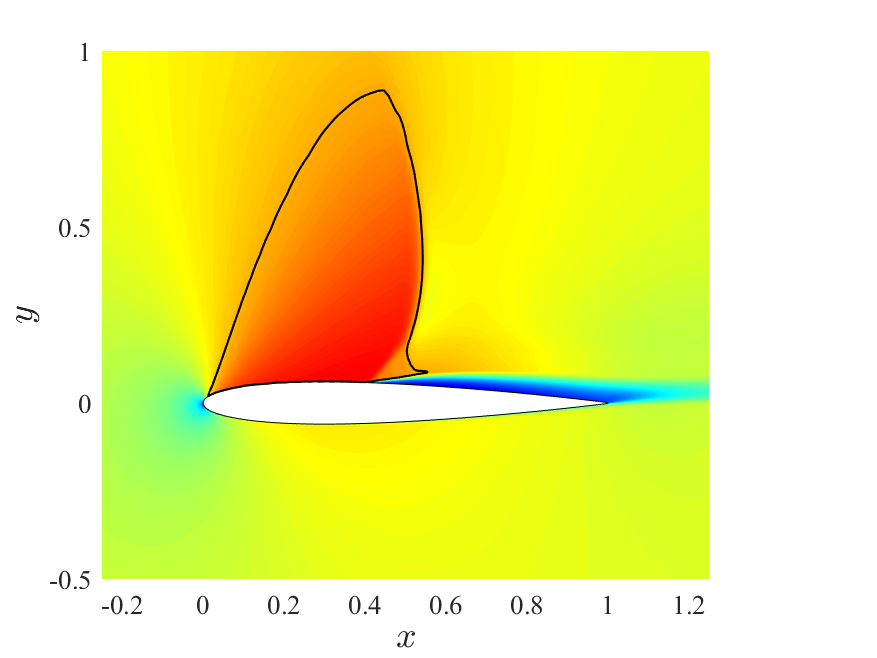}
        \includegraphics[trim={0.2cm 0.15cm 1.25cm 0.7cm}, clip,width=0.425\textwidth]{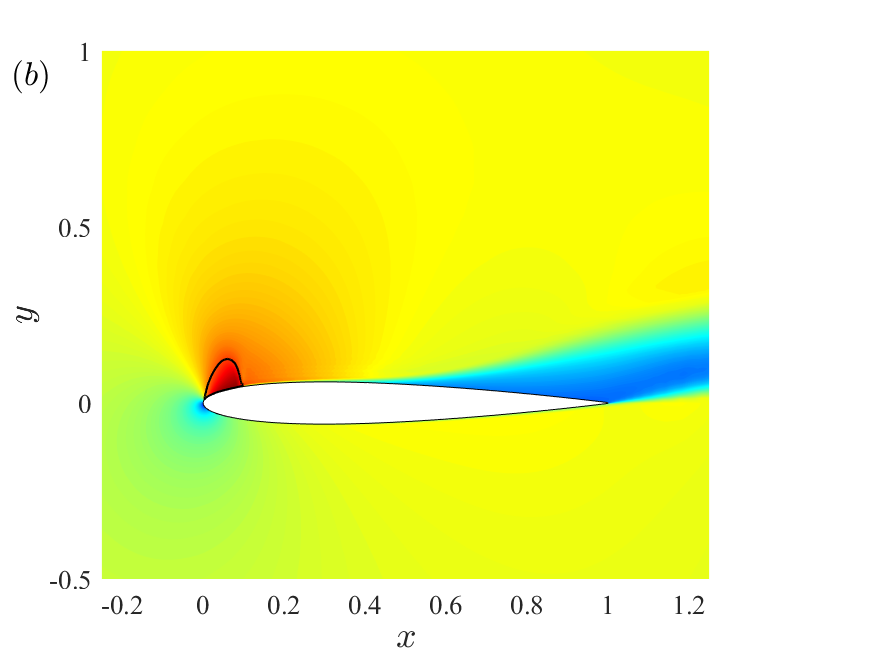}
        \includegraphics[trim={0.2cm 0.15cm 1.25cm 0.7cm}, clip,width=0.425\textwidth]{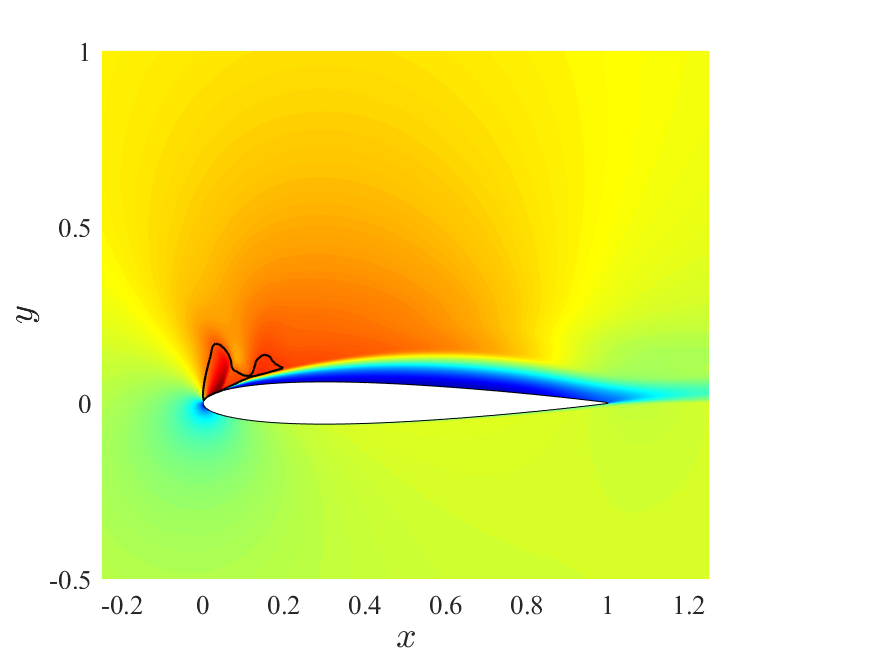}
        \includegraphics[trim={0.2cm 0.15cm 1.25cm 0.7cm}, clip,width=0.425\textwidth]{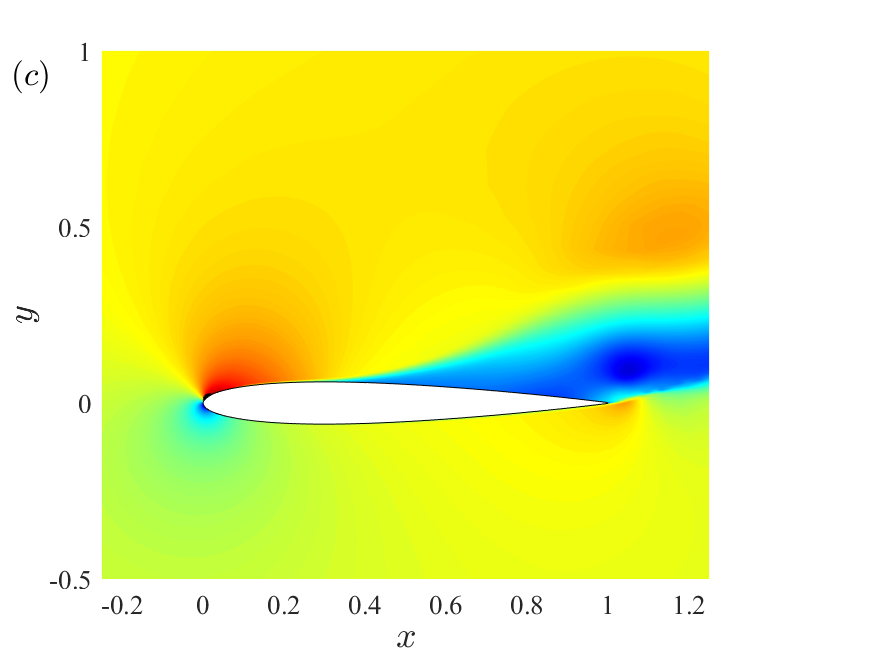}
        \includegraphics[trim={0.2cm 0.15cm 1.25cm 0.7cm}, clip,width=0.425\textwidth]{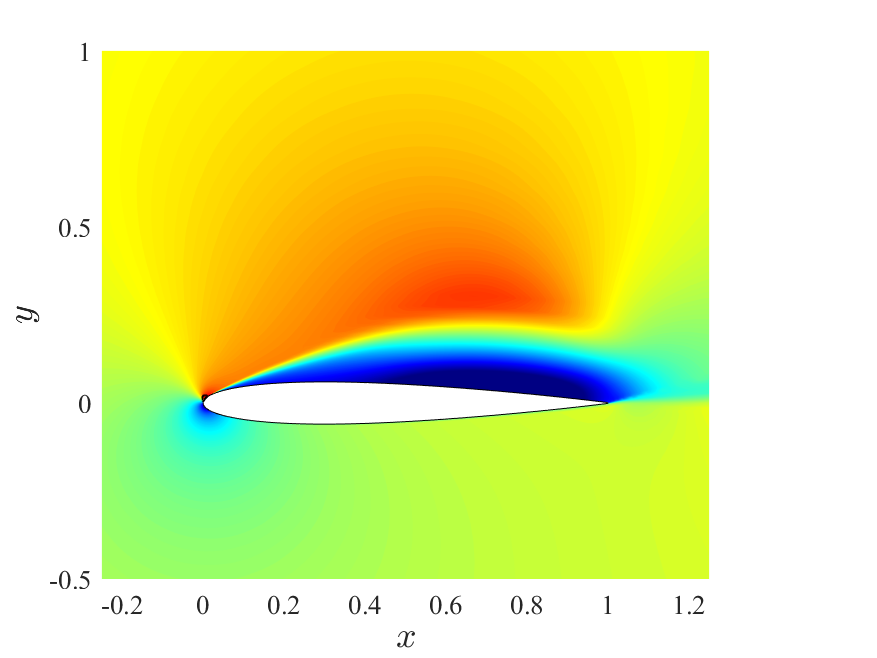}
        \includegraphics[trim={0.2cm 0.15cm 1.25cm 0.7cm}, clip,width=0.425\textwidth]{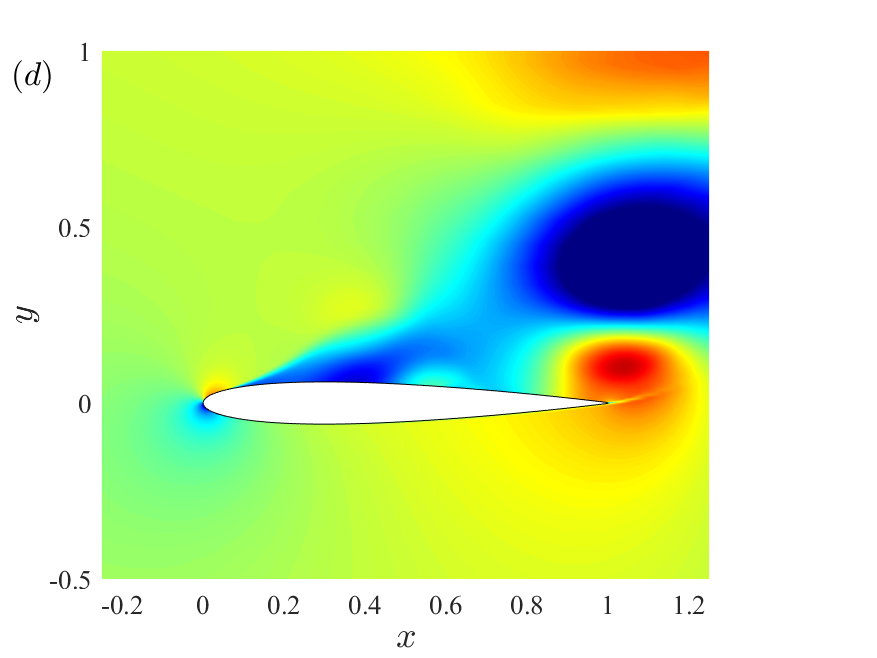}
        \includegraphics[trim={0.2cm 0.15cm 1.25cm 0.7cm}, clip,width=0.425\textwidth]{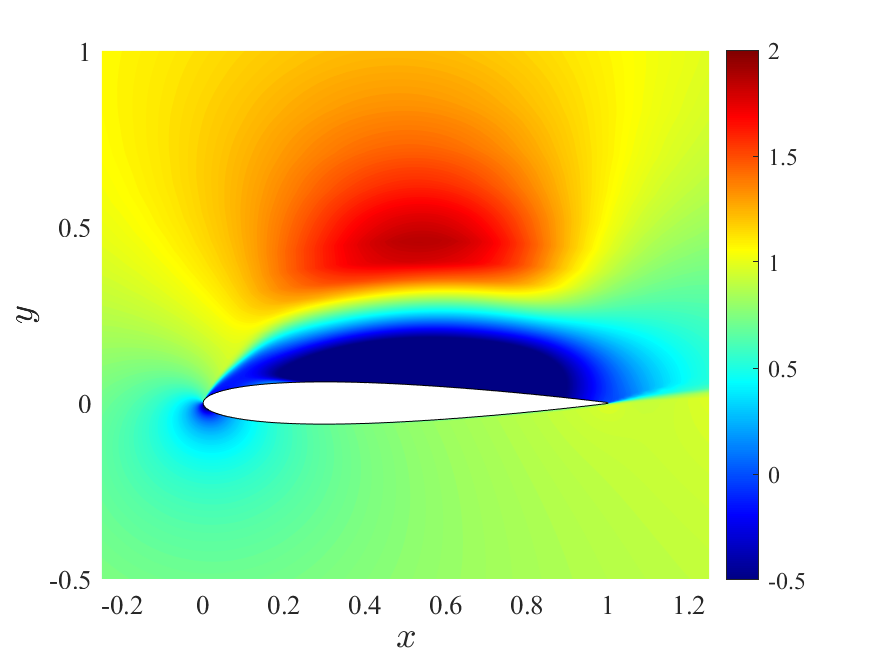}
    \caption{Contours of the dimensionless chordwise velocity component for low-lift (left column) and high-lift phases (right column) for different \(\alpha, M\): (a)  \(4^\circ, 0.76\), (b) \(8^\circ, 0.6\), (c) \(12^\circ, 0.42\) and (d) \(16^\circ, 0.33\). The sonic line is in black.}
    \label{fig:BUffetconts}
\end{figure}

Contours of the chordwise velocity component at approximate high- and low-lift phases of the oscillation cycle observed in Fig.~\ref{fig:Buffet_ClPSD}(a) are shown for different cases in Fig.~\ref{fig:BUffetconts}. Here, the velocity is dimensionless (based on the freestream velocity scale), and the range is fixed manually as -0.5 to 2 to facilitate comparison across cases. Additionally, the sonic line is highlighted using a black curve and indicates the presence of supersonic regions in the flow field. At $\alpha = 4^\circ$ and $M = 0.76$ (Fig.~\ref{fig:BUffetconts}(a)), a large supersonic pocket is observed, with a single shock wave present. The shock wave oscillates periodically in the chordwise direction, accompanied by strong separation and reattachment of the boundary layer, as is typical of transonic buffet\footnote{Supplementary Movie 1, see \url{https://www.youtube.com/playlist?list=PL6tOYmEBo5VN0nsJZ4LJa2WyhsVKw_lp5
}}. For $\alpha = 8^\circ$ and $M = 0.6$ (Fig.~\ref{fig:BUffetconts}(b)), the supersonic region is observed to be relatively small, although present in both the high and low-lift phases. In the high-lift case, in addition to the main supersonic region, a supersonic tongue is seen immediately downstream of it (see Supplementary Movie 2). At higher angles of attack of $\alpha = 12^\circ$ and $16^\circ$ (Fig.~\ref{fig:BUffetconts}(c) and Fig.~\ref{fig:BUffetconts}(d)), the flow field is almost entirely subsonic, although the amplitude of the lift oscillations is of the same order as transonic conditions (see also, Movie 3 and Movie 4). These oscillations are observed for all angles of attack considered from 4$^\circ$ to $19^\circ$ in increments of $1^\circ$, and their Strouhal numbers are of the same order for all cases (see Table~\ref{tab:cases}). This suggests that these oscillations are connected and that they can occur irrespective of whether shock waves exist or are absent and whether the flow is transonic or subsonic, which is consistent with the observations made at low Reynolds numbers \cite{Moise2024}. This hypothesis is further corroborated by the results from incompressible simulations and SPOD in the following sections. 

\begin{figure}[t]
    \centering
        \includegraphics[trim={0.2cm 0.1cm 0.75cm 0.25cm}, clip,width=0.475\textwidth]{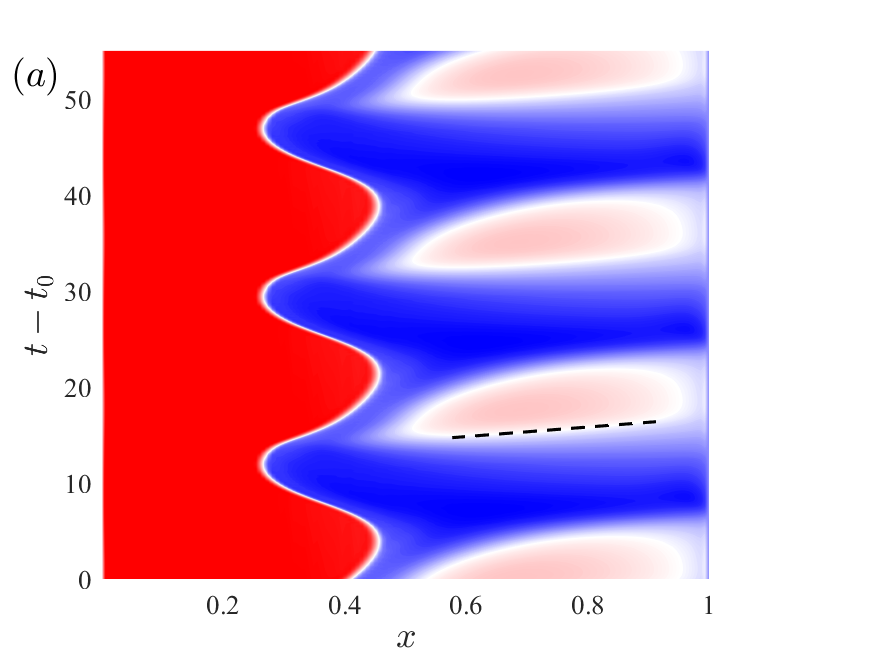}
        \includegraphics[trim={0.2cm 0.1cm 0.75cm 0.25cm}, clip,width=0.475\textwidth]{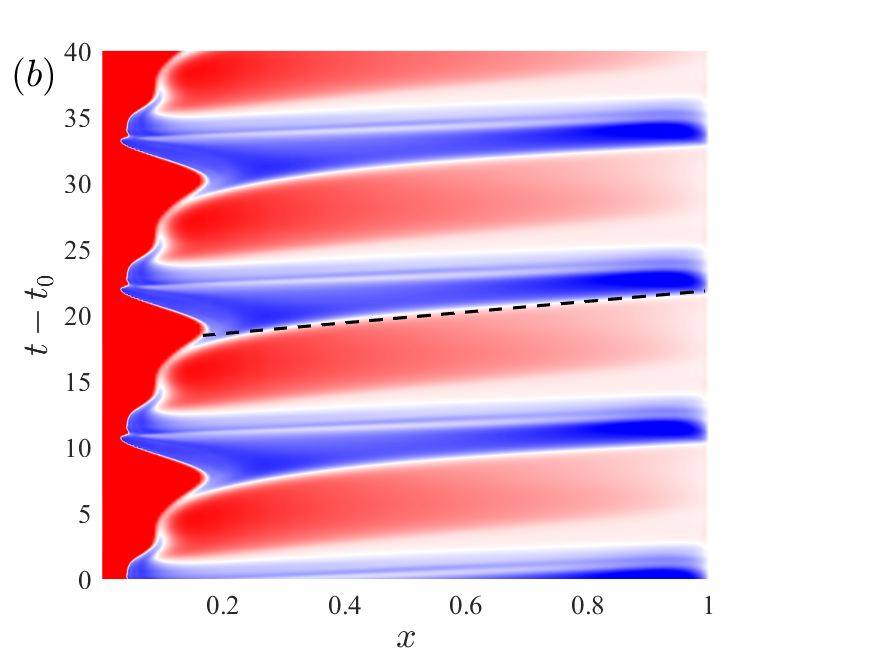}
        \includegraphics[trim={0.2cm 0.1cm 0.75cm 0.25cm},clip,width=0.475\textwidth]{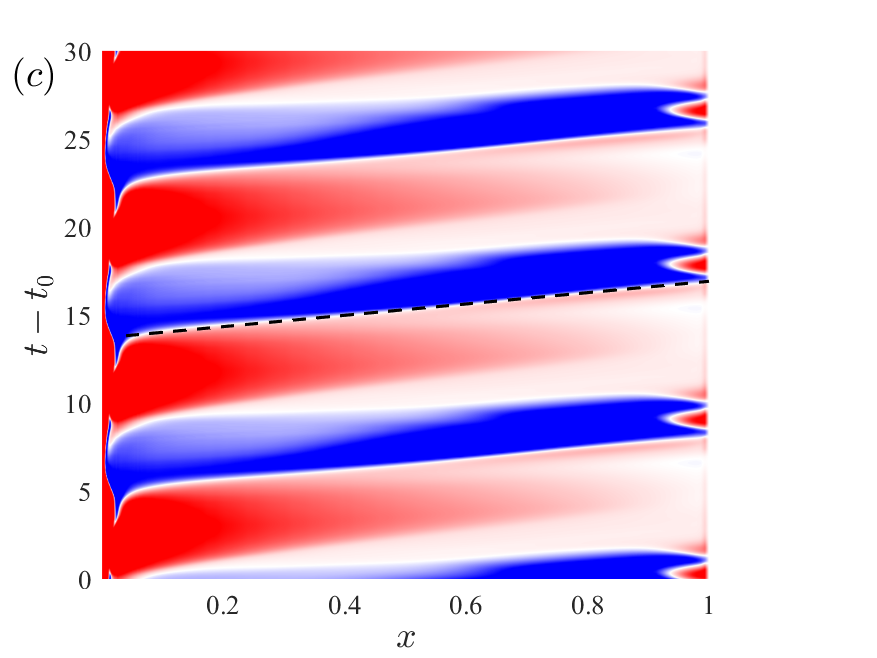}
        \includegraphics[trim={0.2cm 0.1cm 0.75cm 0.25cm},clip,width=0.475\textwidth]{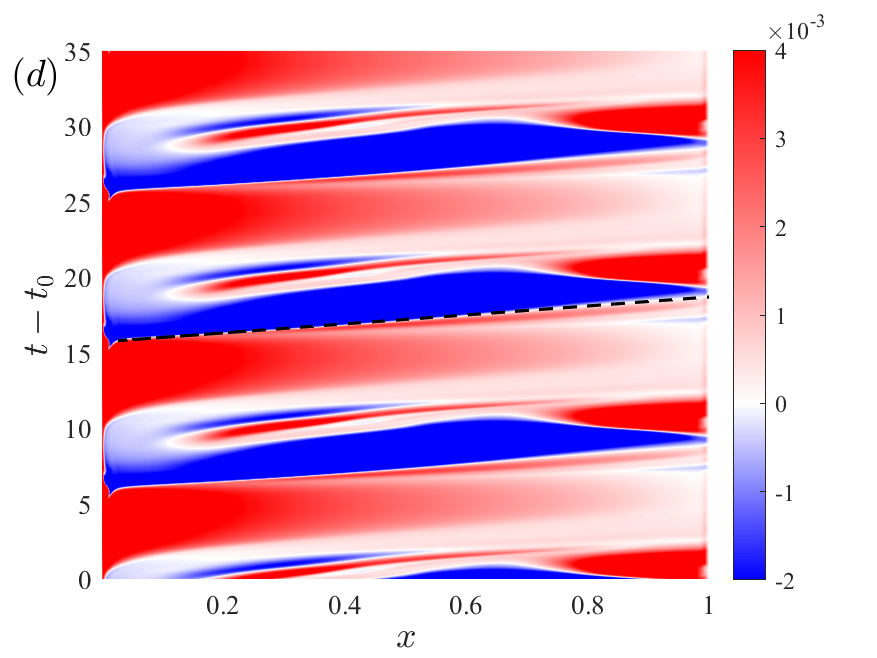}
    \caption{Spatio-temporal evolution of $C_f$ on airfoil's suction surface for different \(\alpha, M\): (a) \(4^\circ, 0.76\), (b) \(8^\circ, 0.6\), (c) \(12^\circ, 0.42\), (d) \(16^\circ, 0.33\).}
    \label{fig:XT_buffet}
\end{figure}

The spatio-temporal evolution of the skin friction coefficient on the airfoil's suction surface is plotted in Fig.~\ref{fig:XT_buffet} for the aforementioned cases. In these plots, white regions correspond to locations where  $C_f \approx 0$, including points of flow separation or reattachment, while red and blue regions represent areas of streamwise flow and flow reversal, respectively. Footprints of structures that propagate downstream are visible in the skin-friction plots, as highlighted by the dashed lines. These are discussed further in Sec.~\ref{subSecFeedback}. For the transonic flow at $\alpha = 4^\circ$ and $M = 0.76$ (Fig.~\ref{fig:XT_buffet}(a)), the boundary layer separates at the foot of the oscillating shock wave at all times (indicated by the white sinusoidal edge of the red region, approximately within the range $0.3 \leq x \leq 0.4$). It can be inferred from the figure that during periods of upstream shock wave motion, the flow downstream of the shock foot remains separated up to the trailing edge. At other times, a separation bubble exists, with the flow reattaching downstream to the shock wave. At $\alpha = 8^\circ$ and $M = 0.6$ (Fig.~\ref{fig:XT_buffet}(b)), similar oscillatory features are observed, but for this case, there are periods in the oscillation cycle when the boundary layer remains fully attached. For higher angles of attack (Fig.~\ref{fig:XT_buffet}(c) and Fig.~\ref{fig:XT_buffet}(d)), the separation location moves up to the leading edge at some time instants, while the flow remains fully attached at other time instants. For all cases, it should be noted that the separation location, when it exists, is dynamically changing over a large extent along the chord and is not localized to the leading or trailing edge (i.e., not a leading edge or trailing edge stall). Although the extent of separation/reattachment is different for the cases shown, the low frequency of the oscillation of the separation/reattachment points remains a defining characteristic of these flows.  

\begin{figure}[t]
    \centering
        \includegraphics[trim={0cm 0cm 0cm 0cm}, clip,width=0.49\textwidth]{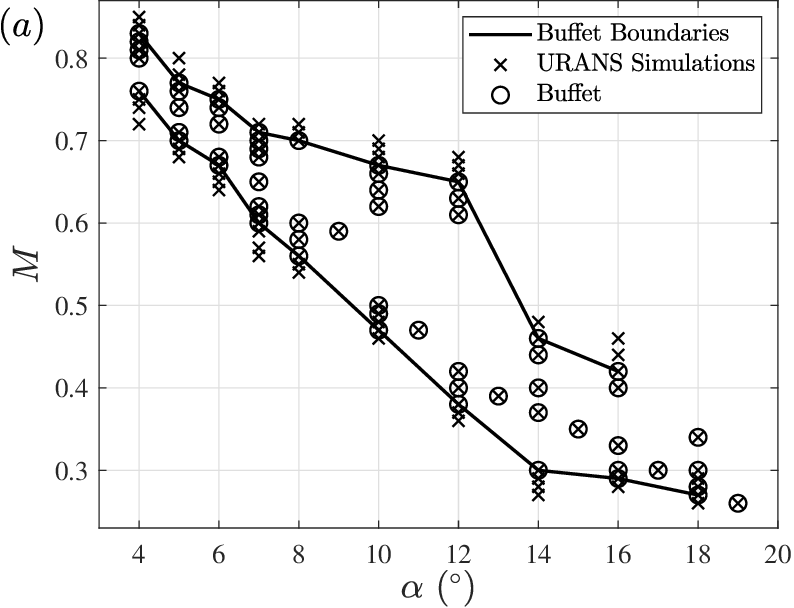}
       \includegraphics[trim={0cm 0cm 0cm 0cm}, clip,width=0.49\textwidth]{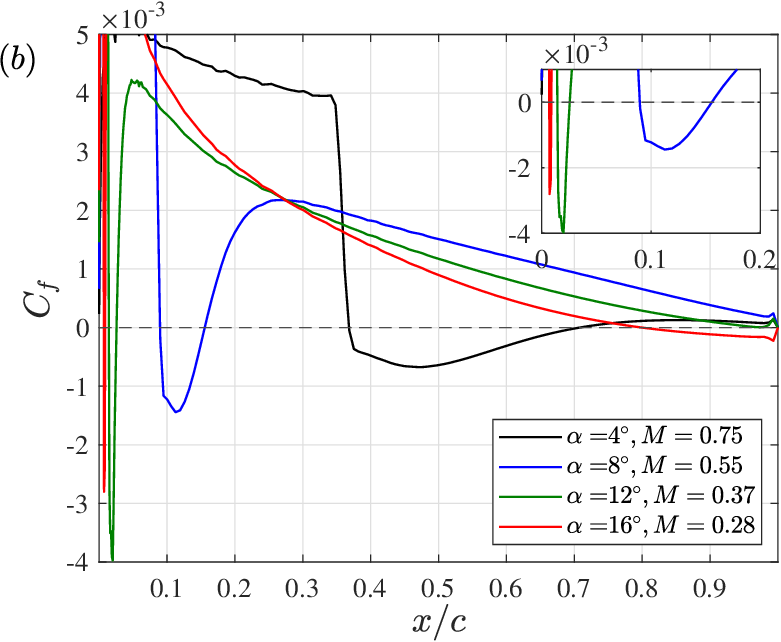}
    \caption{(a) A map of all compressible flow conditions simulated (crosses), with cases involving buffet highlighted using circles. The solid lines mark the onset and offset boundaries. (b) Coefficient of Skin-friction plotted for cases before the onset $M$ for different $\alpha$}
    \label{fig:OnOffset}
\end{figure}

A map of all conditions where buffet occurs in compressible URANS simulations is plotted in Fig.~\ref{fig:OnOffset}(a). The circles highlight situations where sustained buffet is observed. The solid lines delineate the onset and offset boundaries of these oscillations. These boundaries are determined by the amplitude of the lift oscillations, with pre-onset conditions having $\Delta C_L/2 =(\max{(C_L)}-\min(C_L))/2 < 10^{-2}$, while post-onset, $\Delta C_L/2 > 0.1$ for all cases examined (not shown). This indicates the abrupt growth of these oscillations, with a change of $\Delta M = 0.01$ inducing lift oscillation amplitudes of $\Delta C_L = 0.1$. For  $4^\circ\leq\alpha\leq16^\circ$, buffet was observed in a narrow range of $M$ between the onset and offset boundaries. For $17^\circ\leq\alpha \leq 19^\circ$, although regular oscillations were observed for some $M$ (as highlighted in Fig.~\ref{fig:OnOffset}(a) using symbols), as noted in Sec.~\ref{subSecMethodTurbModels}, the oscillations at other conditions were found to be intermittent (see Appendix, Fig.~\ref{figApp:LongTime}(a)). These intermittent oscillations still exhibit a discrete peak in the power spectrum of lift, which is related to buffet (similar to that shown in Fig.~\ref{fig:LFO_Cl_PSD}). At other conditions, these oscillations led to numerical instabilities that caused divergence issues, and thus, it was not feasible to determine an offset boundary for these $\alpha$. The origin of this issue is unclear -- it might arise due to hysteresis and bistable features reported for these states at high $Re$ (e.g., \cite{Hristov2018,Busquet2021}, or it might be due to the turbulence models used (simulations using SA model do not exhibit such intermittent behaviour), or numerical instability, but this is not explored further. For $\alpha \geq 20^\circ$, regular oscillations were not observed for any $M$ attempted, and thus, we have limited our search for buffet in the compressible simulations to $\alpha \leq 19^\circ$.

For $4^\circ\leq \alpha \leq 14^\circ$, it appears that there is an approximately linear relation between the onset $M$ and $\alpha$, with the slope of the onset boundary being approximately -0.04. To explore for possible signatures that can be used to predict buffet onset, we examine the flow features for the highest $M$ below the onset value (i.e., pre-onset, steady situation) at different $\alpha$. The chordwise variation of $C_f$ for these pre-onset conditions is shown in Fig.~\ref{fig:OnOffset}(b) for different $\alpha$. It is seen that under all conditions, there is a separation bubble present (i.e., a finite pocket of $C_f < 0$ within $0\leq x/c\leq 1$). This indicates that buffet occurs in a narrow range between conditions where the flow has a steady separation bubble and where the flow is fully stalled, irrespective of the flow being transonic or subsonic. Note that the presence of a separation bubble of an extremely small streamwise extent is commonly reported in experiments as preceding the onset of leading-edge stall (e.g., \cite{McCullough1951, Aniffa2023a}), but it is important to note that once oscillations set in, the instantaneous results from the present study show that it is not ideal to classify the flow as exhibiting leading edge stall. This is because the separation point is found to traverse over a large extent of the airfoil, and not localized only at the leading edge at different times  (cf. Fig.~\ref{fig:XT_buffet}d).

In summary, all of the above results indicate that oscillations of a low frequency ($St \sim O(10^{-1}$)) can be sustained at all angles of attack in the range of $4^\circ$ to $19^\circ$ by appropriately changing the freestream Mach number. At low angles of attack, the flow is transonic, involving large-scale shock wave motion (\textit{i.e}, transonic buffet), and at high angles, the flow field is entirely subsonic and approximately incompressible (\textit{i.e.}, LFO). The spatial structure of the flow oscillations at high- and low-lift phases and the spatio-temporal characteristics of separation and reattachment of the boundary layer on the suction side are also seen to have qualitative similarities for the entire range of parameters studied. Buffet onset occurs at decreasing $M$ for increasing $\alpha$ (linear relation), and a steady separation bubble of a finite chordwise extent is present just prior to the onset of buffet in all regimes.

\section{Incompressible low-frequency oscillations}
\label{secIncompressibleRANS}

In the preceding section, buffet-like oscillations have been shown to occur at freestream Mach numbers as low as $M = 0.26$ at $\alpha = 19^\circ$. In this section, we further explore whether such oscillations can be sustained when simulating the incompressible RANS equations at the same Reynolds number of $Re = 10^7$. In the limit of incompressibility, we have the scaled speed of sound, $a_\infty/U_\infty\rightarrow \infty$, implying $M\rightarrow 0$. Thus, LFO is expected at a higher $\alpha$ for the incompressible regime than $\alpha = 19^\circ$ at which it is observed for $M = 0.26$, as suggested by the trends seen in Fig.~\ref{fig:OnOffset}(a). As noted in Sec.~\ref{subSecMethodTurbModels}, the results are sensitive to the turbulence model used and thus, we report results for both the SA and SST $k-\omega$ model. For the former case, simulations based on the incompressible RANS equations were carried out for $\alpha = 19^\circ$ and $20^\circ \le \alpha < 22^\circ$ in steps of $0.25^\circ$. Oscillations were observed in a narrow range of $20.5^\circ \leq\alpha\leq 21.5^\circ$, which exhibit the same characteristics as incompressible LFO. Narrow ranges for LFO have been observed in similar settings in experiments, with Atallah \textit{et al.} \cite{Atallah2024} noting that the onset and offset angle range could be as small as $\Delta\alpha = 0.18^\circ$ for the NACA0012 three-dimensional wing at $Re\approx 10^5$ and natural transition conditions. We choose the reference case of $\alpha = 21^\circ$ for further examination. For the latter case, the flow becomes unsteady for $\alpha \geq 22^\circ$, but the oscillations are intermittent, and divergence issues were observed at higher incidence angles. The reference case is chosen as $\alpha = 22^\circ$ for this turbulence model.

\begin{figure}[t]
    \centering
        \includegraphics[trim={0cm 0cm 0cm 0cm}, clip,width=0.47\textwidth]{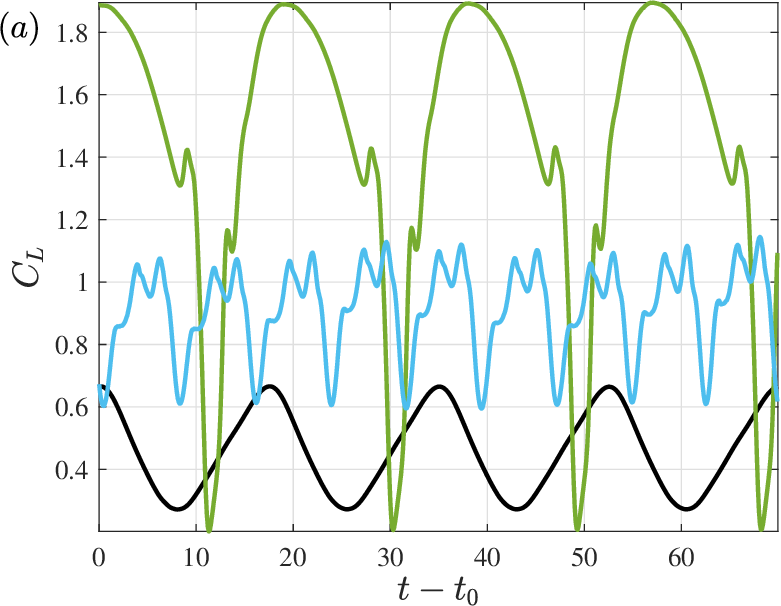}
        \hspace{0.5cm}
        \includegraphics[trim={0cm 0cm 0cm 0cm}, clip,width=0.47\textwidth]{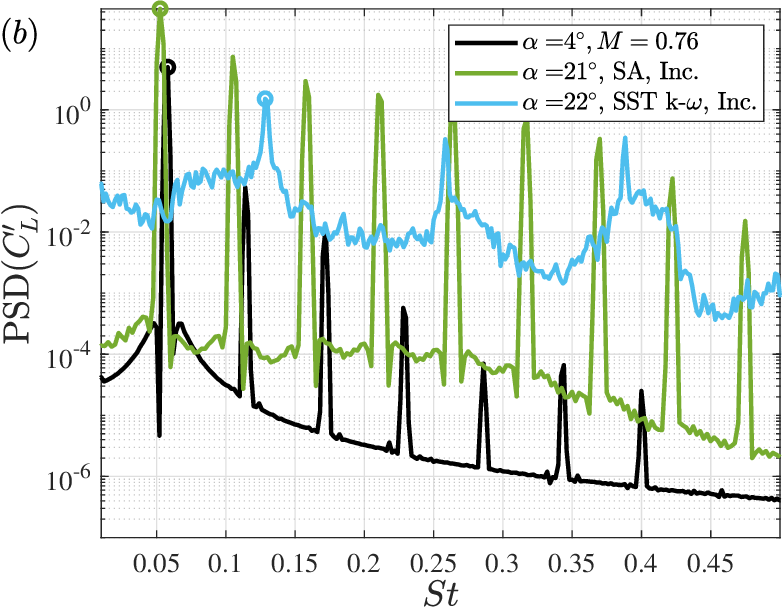}
    \caption{(a) Temporal variation of the lift coefficient past transients and (b) the corresponding power spectrum of its fluctuating component obtained from incompressible and compressible simulations of NACA 0012 at \(Re = 10^7\).}
    \label{fig:LFO_Cl_PSD}
\end{figure}

The temporal variation of the lift coefficient from the two references incompressible cases is compared with buffet from compressible RANS simulations in Fig.~\ref{fig:LFO_Cl_PSD}(a). Sinusoidal oscillations at a low frequency are observed for the cases compared. The corresponding PSD spectra of the fluctuating component are compared in Fig.~\ref{fig:LFO_Cl_PSD}(b) with peaks in the spectra associated with buffet and incompressible LFO highlighted using circles. Thus, sustained oscillations at a low frequency ($St \sim O(10^{-1})$) are present for incompressible conditions, which is of the same order as that seen in the compressible RANS simulations, including those at transonic conditions. It is emphasized that although the peak for the case of SST $k-\omega$ occurs at $St_b \approx 0.12$ which is higher than that for the other two cases compared ($St_b \approx 0.05$), it is still within the range observed in the compressible simulations ($0.05 \leq St_b \leq 0.15$), and the differences seen could be due to the sensitivity of $St_b$ to $\alpha$ and $M$.

\begin{figure}[t]
    \centering
        \includegraphics[trim={0.2cm 0.15cm 1.25cm 0.7cm}, clip,width=0.45\textwidth]{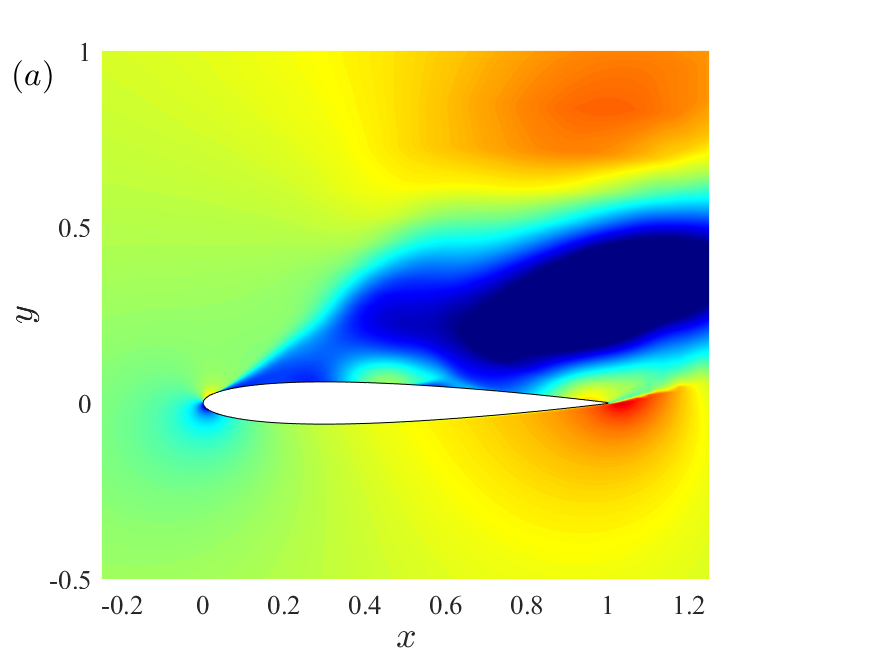}
        \includegraphics[trim={0.2cm 0.15cm 1.25cm 0.7cm}, clip,width=0.45\textwidth]{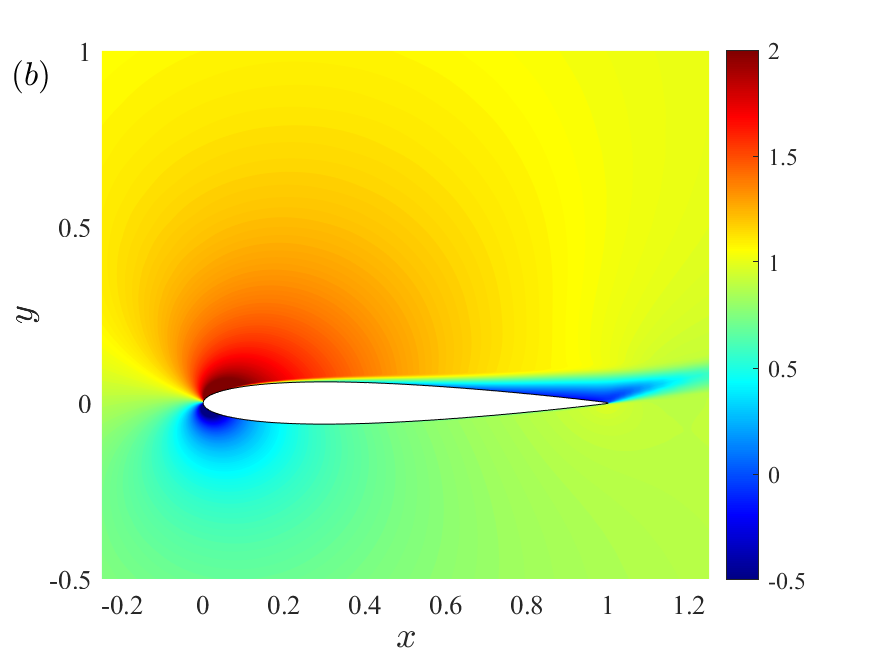}
        \includegraphics[trim={0.2cm 0.15cm 1.25cm 0.7cm}, clip,width=0.45\textwidth]{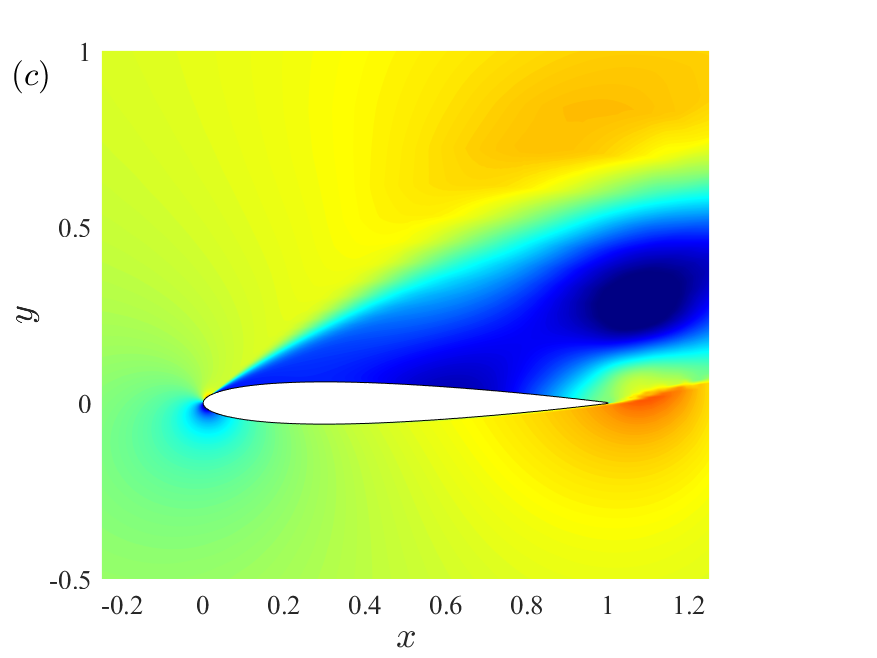}
        \includegraphics[trim={0.2cm 0.15cm 1.25cm 0.7cm}, clip,width=0.45\textwidth]{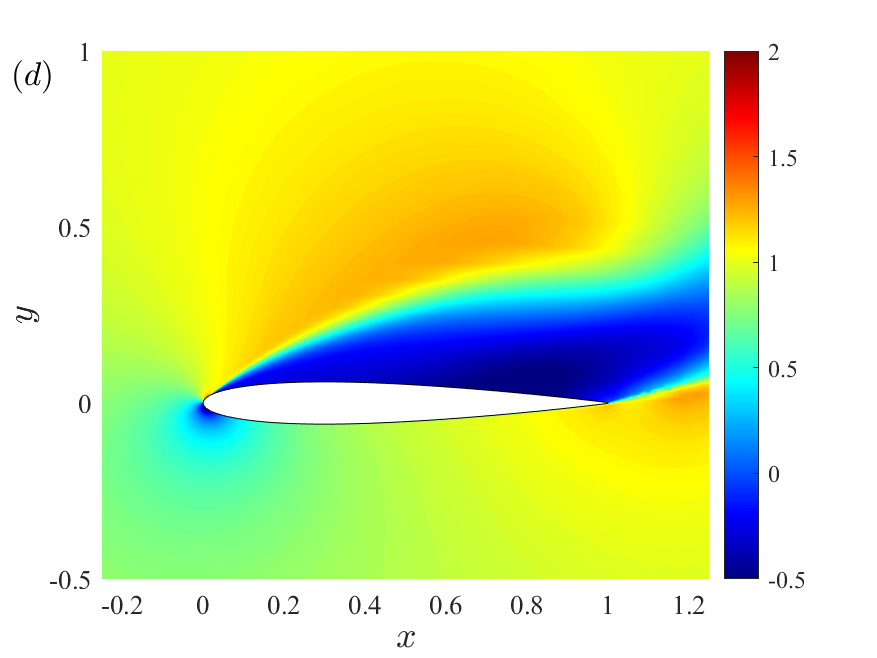}
    \caption{Contours of the dimensionless chordwise velocity for low-lift (left column) and high-lift phases (right column) for (a-b) \(\alpha\) = \(21^\circ\), SA model and (c-d) \(\alpha = 22^\circ\) SST \(k-\omega\).}
    \label{fig:LFO_contours}
\end{figure}

\begin{figure}[t]
    \centering
        \includegraphics[width=0.49\textwidth]{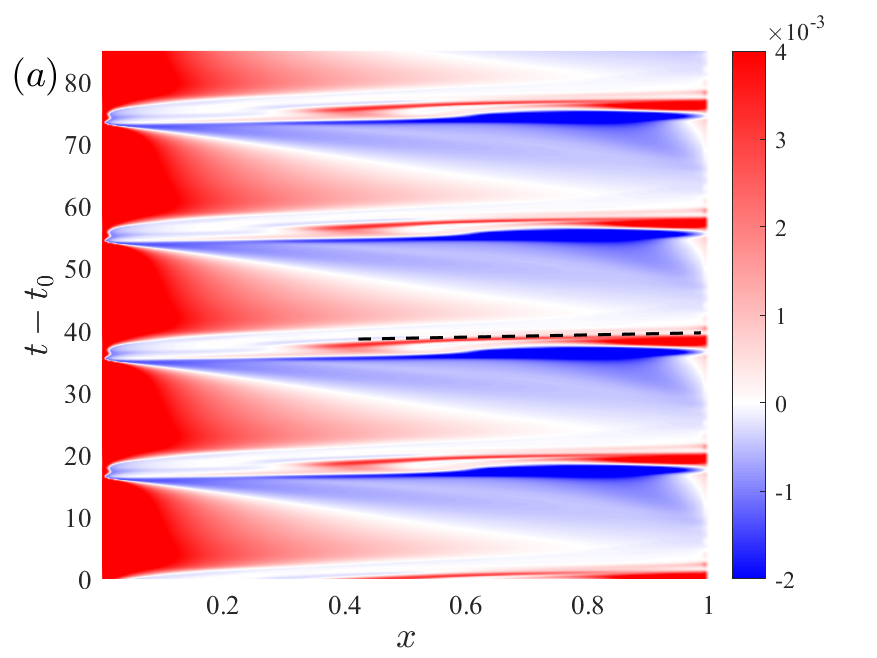}
        \includegraphics[width=0.49\textwidth]{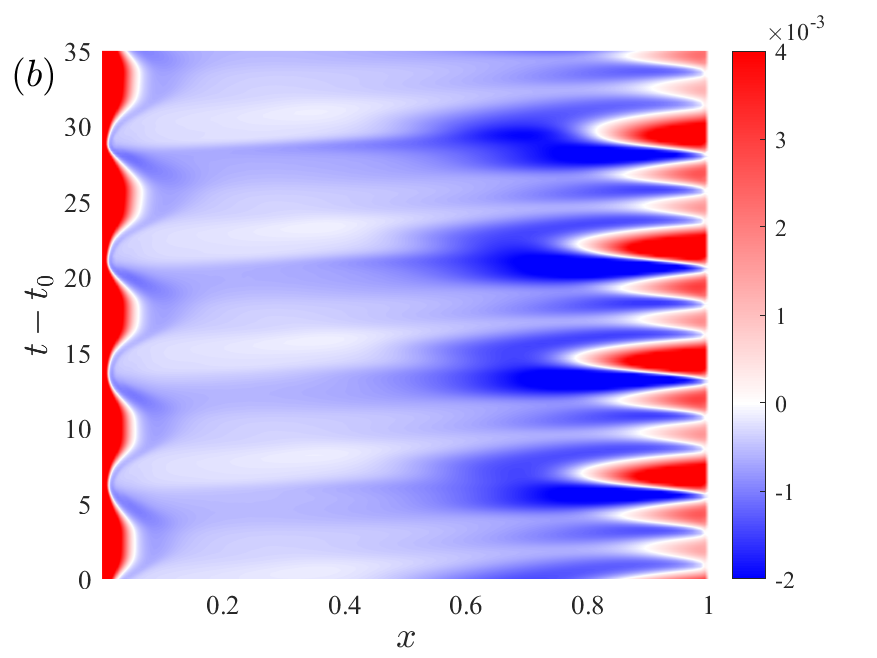}

       \caption{Spatio-temporal evolution of skin friction coefficient for (a) \(\alpha\) = \(21^\circ\), SA model and (b) \(\alpha\) = \(22^\circ\) SST \(k-\omega\), for incompressible low-frequency oscillations.}
    \label{fig:xtLFO}
\end{figure}

Contours of the scaled chordwise velocity component at high- and low-lift phases of the LFO cycle are shown in Fig.~\ref{fig:LFO_contours} for the two reference incompressible cases. Similar topological characteristics to those observed in the compressible simulations are seen here for the SA model, with the low-lift phase involving a large separation region and the high-lift phase involving a relatively attached flow (cf. Fig.~\ref{fig:BUffetconts}). However, for the case of the SST $k-\omega$ model, the flow remains mostly separated at both high and low-lift phases, suggesting that the flow is in the post-stall regime. The separation characteristics are further scrutinized by examining the spatiotemporal variation of the skin-friction coefficient on the airfoil's suction side in Fig.~\ref{fig:xtLFO}. For the case of the SA model, the oscillations lead to periodic switching between stalled and un-stalled states, as is typical of incompressible LFO \cite{Zaman1989}, with large-scale variations in the chordwise locations of the separation/reattachment points qualitatively resembling the behaviors observed in the compressible regime (Fig.~\ref{fig:XT_buffet}). In addition to the periodic oscillation of the separation point (i.e., white boundary) that traverses almost the entire chord in one oscillation cycle, there are also what appear to be structures that propagate downstream (i.e., positive slope of $\approx 0.5$ in the $x-t$ diagram) as highlighted by the dashed line. These structures are also observed in compressible simulations at all conditions and will be discussed in the context of the feedback-loop models in Sec.~\ref{subSecFeedback}. For the case of the SST $k-\omega$ model, the flow is separated over most of the airfoil at all times, but the reverse flow magnitude varies with time regularly, and the separation point is seen to move sinusoidally in the fore-aft direction, albeit only in the vicinity of the leading edge. It is not evident if the oscillations are related to LFO, as the flow does not reattach at some instants in the oscillation cycle. Nevertheless, the spatial structure of the oscillations is similar to that observed for all other cases, as shown later in Sec.~\ref{secSPOD} (see Fig.~\ref{fig:Buffet_SPODmodes}). Similar behavior is also reported in the transonic regime for high freestream Mach numbers, where buffet oscillations continue to exist even when the boundary layer is separated over most of the airfoil at all times (see case of $M=0.85$ in Ref.~\cite{Moise2022}). 

In summary, the results presented in this section further indicate that the flow oscillations on airfoils are not unique to the transonic regime, but can be sustained at low frequencies even in the fully incompressible regime when the angle of attack is sufficiently high, implying that transonic buffet and incompressible LFO are connected. It is important to note that the results based on the SA turbulence model are consistent with the results from the compressible simulations, but for the SST $k-\omega$ model, there are significant differences in the separation characteristics, with the flow remaining fully stalled at all times, and it is not clear which of these models is relevant and thus, higher-fidelity approaches are needed to clarify these observations. Nevertheless, the signature of buffet, i.e., the low frequency of the oscillations $St_b \sim O(10^{-1})$ is present for both turbulence models, and we will show next that the spatial structure of these incompressible flow oscillations is also similar irrespective of the models, implying that even if the base flow predictions are different, the oscillatory spatio-temporal characteristics are consistent and resemble that observed in the transonic regime.

\section{Spectral proper orthogonal decomposition}
\label{secSPOD}

\begin{figure}[t]
    \centering
        \includegraphics[width=0.50\textwidth]{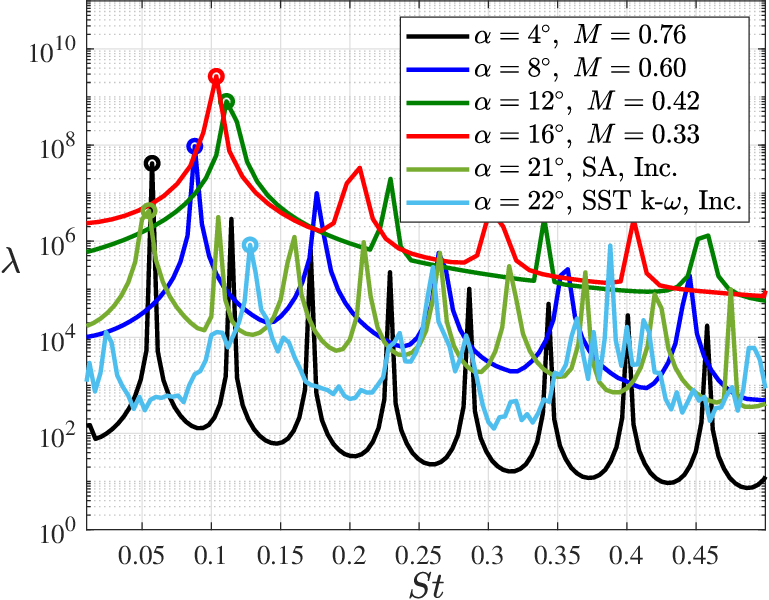}
    \caption{Eigenspectra of the dominant SPOD mode shown for representative cases of NACA 0012 airfoil at \(Re = 10^7\).}
    \label{fig:SPODSpectra}
\end{figure}

The lift oscillations observed for the various cases simulated here all have $St_b \sim O(10^{-1})$, suggesting a similarity in temporal features. Here, we further investigate similarities in the spatio-temporal characteristics of the oscillatory field by analyzing the URANS results using SPOD, which looks at the features of the global flow field over the entire domain. The eigenvalue spectra of the most energetic SPOD mode are compared for different cases in Fig.~\ref{fig:SPODSpectra}. The spectra resemble those based on the lift coefficient (cf. Fig.~\ref{fig:Buffet_ClPSD} and Fig.~\ref{fig:LFO_Cl_PSD}), with the dominant peaks occurring for the same range of $St_b$ in the SPOD spectra as well (i.e., $0.05\leq St_b \leq 0.15$). Note that for the case of the intermittent oscillations observed for the reference case of the SST $k-\omega$ model used for incompressible simulations, there are low-amplitude peaks at low frequencies of $St_b\approx 0.02$ in addition to the `buffet' mode associated with the dominant peak at $St_b\approx 0.12$ (highlighted by a circle). It is unclear if they represent any physical feature, but the mode's spatial features were similar to the buffet mode's features, suggesting that they may be weak sub-harmonics. This is not explored further, as they are of relatively low energy (eigenvalue, $\lambda$, two orders lower) compared to the buffet mode.
The spatial structure of all the buffet modes associated with the dominant peaks in the SPOD spectra (highlighted by circles) is examined next. A visual comparison is made before quantifying the spatial similarity between these oscillations. 

\begin{figure}[hbt!]
    \centering        \includegraphics[trim={0.25cm 0cm 1.5cm 0cm}, clip, width=0.48\textwidth]{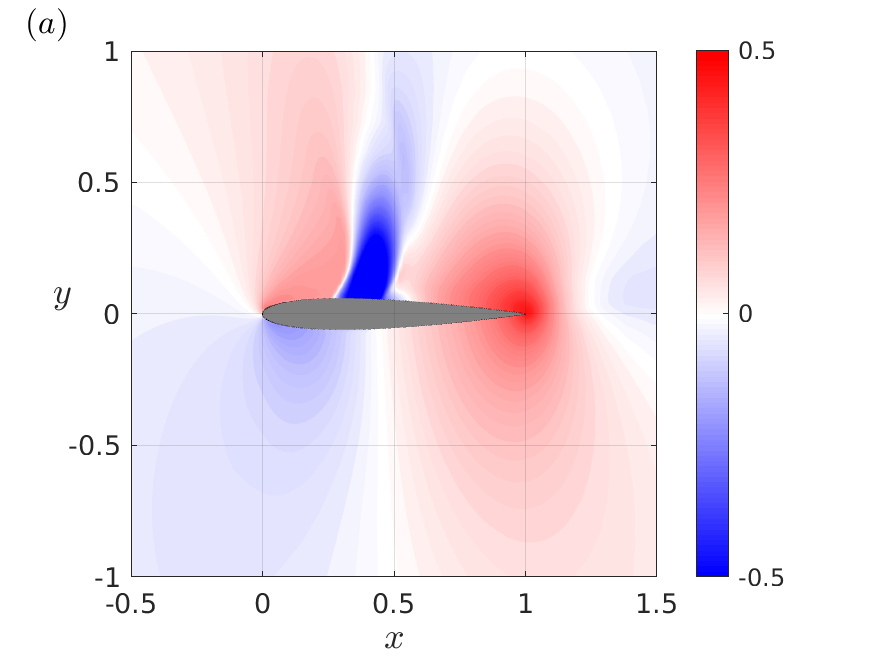}        \includegraphics[trim={0.25cm 0cm 1.5cm 0cm}, clip, width=0.48\textwidth]{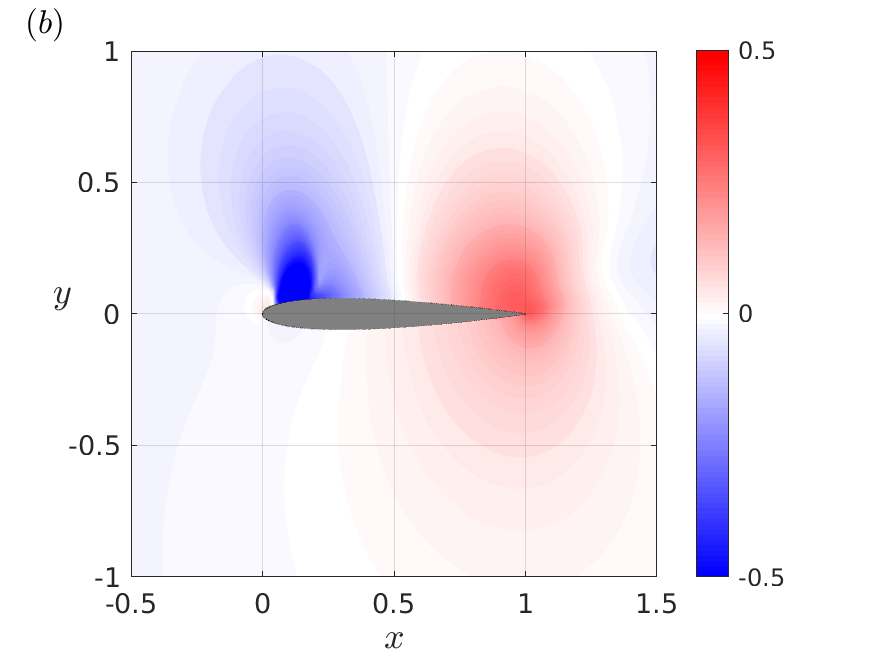} \includegraphics[trim={0.25cm 0cm 1.5cm 0cm}, clip, width=0.48\textwidth]{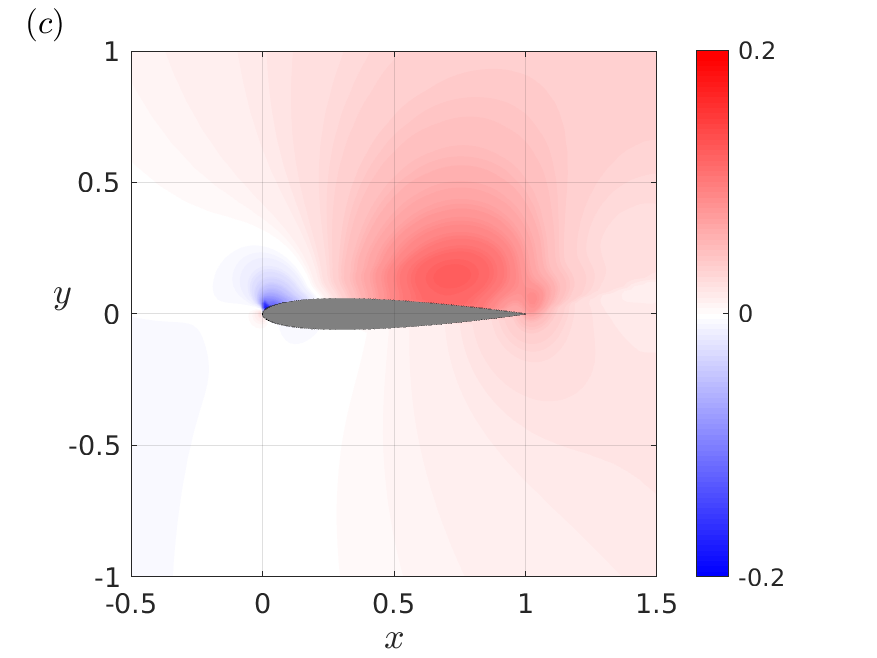}\includegraphics[trim={0.25cm 0cm 1.5cm 0cm}, clip, width=0.48\textwidth]{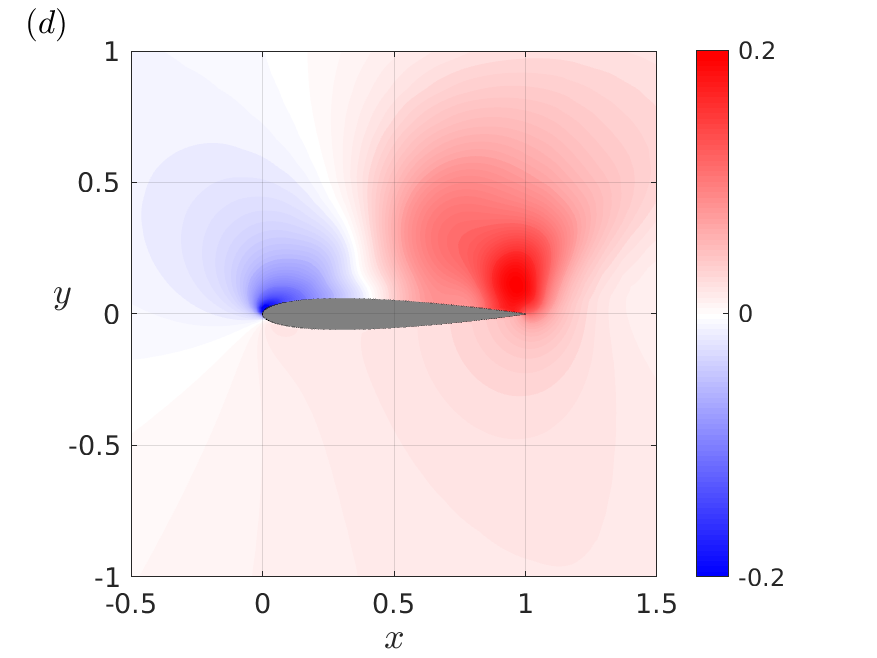}
  \includegraphics[trim={0.25cm 0cm 1.5cm 0cm}, clip,width=0.48\textwidth]{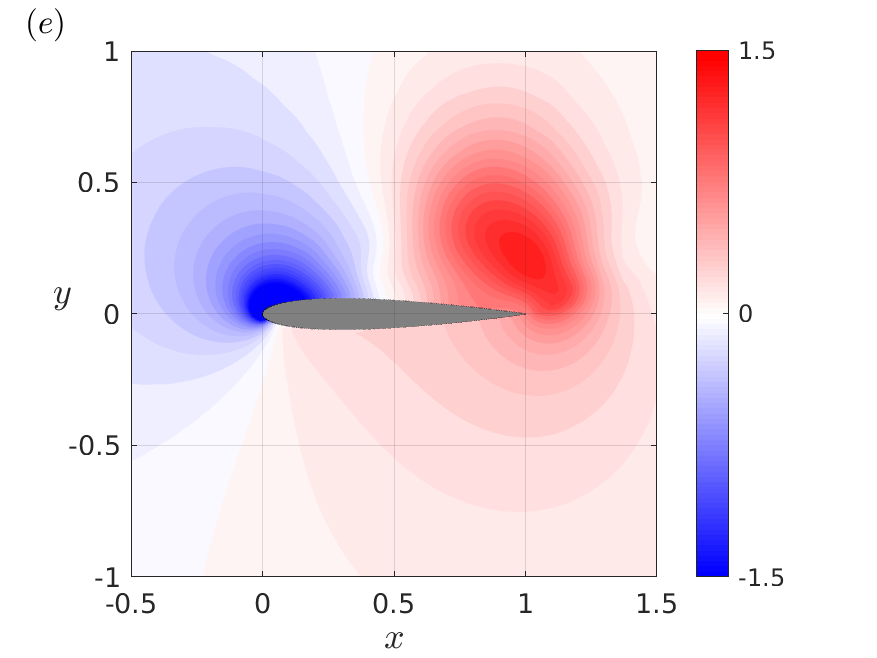}
  \includegraphics[trim={0.25cm 0cm 1.5cm 0cm}, clip,width=0.48\textwidth]{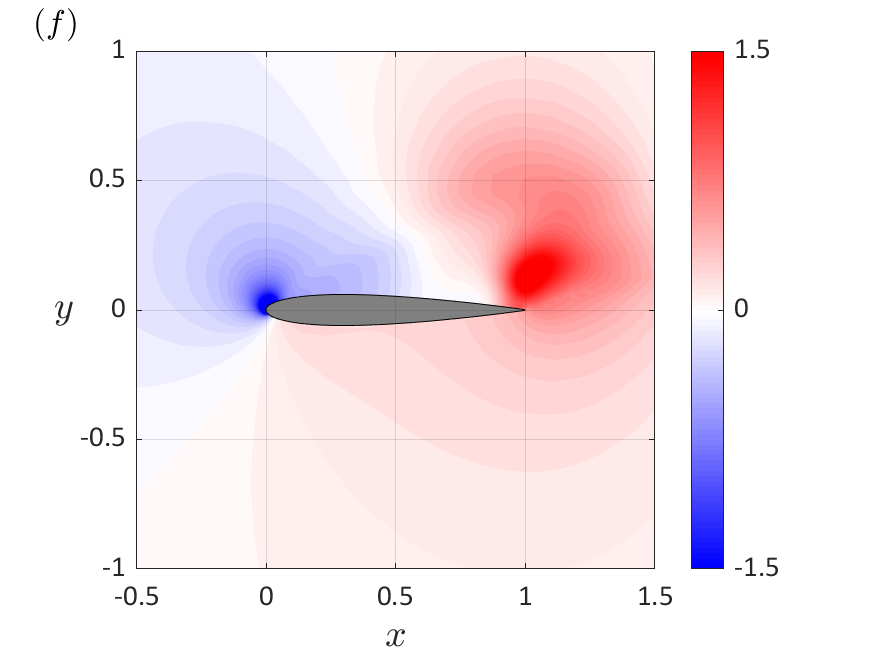}

    \caption{Pressure field from SPOD modes of NACA 0012 airfoil at \(Re = 10^7\) computed using compressible URANS simulations for different \((\alpha, M, St_{b})\): (a) \((4^\circ,0.76,0.06)\), (b) \((8^\circ,0.60,0.09)\), (c) \((12^\circ,0.42,0.11)\), (d) \((16^\circ,0.33,0.10)\), incompressible URANS simulations for (e) SA model at \((\alpha, St_{b}) = (21^\circ,0.06)\), and (f) SST  \(k-\omega\) \((22^\circ,0.12)\).}
    \label{fig:Buffet_SPODmodes}
\end{figure}

The pressure fields from SPOD modes corresponding to the buffet frequency at an arbitrary phase in their oscillatory cycle are shown in Fig.~\ref{fig:Buffet_SPODmodes}. The typical case of transonic buffet ($\alpha = 4^\circ$ and $M = 0.76$) is shown in Fig.~\ref{fig:Buffet_SPODmodes}(a), for which the shock wave on the suction side oscillates in the fore-aft direction. For this case, the pressure fluctuation field in the vicinity of the shock wave, \textit{i.e.}, the blue region centered at $x\approx 0.5$ on the suction side, is out of phase with that at the trailing edge, \textit{i.e.}, the red region, at some instant in the oscillatory cycle. This qualitative feature is typical of transonic buffet and has been reported in SPOD modes for a wide range of transonic flow conditions on various airfoils at both low and high Reynolds numbers \cite{Moise2022, Moise2022Trip, Moise2024, Zauner2024}. The blue region associated with the shock wave motion can also be interpreted as the region related to the separation point oscillation, implying an out-of-phase relation between the pressure fluctuation in the vicinity of the separation point and the trailing edge. As $\alpha$ is increased and $M$ is reduced, we observe that this blue region, observed originally at mid-chord on the suction side, now shifts towards the leading edge (Fig.~\ref{fig:Buffet_SPODmodes}(b-d)). Nevertheless, the characteristic feature is preserved, with this region being out of phase with that at the trailing edge. Note that the shock wave is absent at higher angles of attack, but the separation characteristics have visual similarities, and this is reflected in the topological similarity seen in the SPOD modes' spatial structures. This is also observed even when fully ignoring compressibility effects, as seen in Fig.~\ref{fig:Buffet_SPODmodes}(e) and Fig.~\ref{fig:Buffet_SPODmodes}(f), where the SPOD mode associated with LFO from the incompressible RANS simulations is shown. It is important to note that the spatial structure of these buffet modes is qualitatively different from that of other types of oscillations that occur in flows over airfoils. For example, the separation-bubble mode \cite{Dandois2018} that is shown to coexist with transonic buffet by Zauner \textit{et al.} has energy concentrated at the foot of the shock wave (see Fig.~6, p. 1036 in Ref.~\cite{Zauner2020FTaC}), while the wake mode reported in various studies resembles a von K\'arm\'an vortex street with energy concentrated in the wake \cite{Sartor2015, Grossi2014}. It is also interesting to note that although intermittent oscillations occur when the SST $k-\omega$ model is used for the incompressible simulations, the mode associated with the highest energy content in SPOD still has a striking topological similarity to that seen for other cases. This is also seen when compared with the case of the SA turbulence model (Fig.~\ref{fig:Buffet_SPODmodes}(e) and Fig.~\ref{fig:Buffet_SPODmodes}(f)), suggesting that although turbulence models might affect the overall spatio-temporal dynamics of the flow in the incompressible regime at the conditions studied here, the dominant coherent oscillatory feature remains similar.

To quantify the similarity between different SPOD modes obtained, correlation coefficients (see Sec.~\ref{MethodsSPOD}) are calculated between SPOD modes. The cases that are compared are related by an incremental change in the flow parameters. The computed correlations are reported in Table~\ref{correaltion_table}, over the entire range of angles of attack studied, from the transonic regime to incompressible flow. The high degree of correlation ($\geq 50\%$) between all compressible cases gives further quantitative evidence that the subsonic oscillations observed are linked to transonic buffet. Furthermore, the correlation coefficient between the compressible case at $\alpha = 19^\circ$ and the incompressible LFO at $\alpha = 21^\circ$ (SA model) is $61\%$, and $\alpha = 22^\circ$ (SST $k-\omega$ model is $79\%$. Considering the larger differences in the angle of attack, wake orientation, and Mach numbers, the good correlation also suggests that the incompressible and compressible modes are linked and have topological similarity. Thus, the present SPOD results give a stronger quantitative measure to support the hypothesis that transonic buffet and incompressible LFO are linked.

\begin{table}[t]
    \centering
    \caption{Magnitude of Pearson correlation coefficient between pressure modes obtained from SPOD of URANS results.}
    \label{correaltion_table}
\begin{tabular}{c c c c c c c}

 \(\alpha_1 (^\circ)\) & \(\alpha_2 (^\circ)\) & \(M_1\) & \(M_2\) & \(St_{b1}\) & \(St_{b2}\) & Correlation(\%) \\
    4   & 5   & 0.76   & 0.71   & 0.06   & 0.07   & 70.6   \\
    5 & 6 &0.71 & 0.68 & 0.07& 0.07& 81.0\\
    6 & 7 &0.68& 0.65&0.07&0.08&78.3\\
    7 & 8 &0.65&0.6&0.08&0.09&83.4\\
    8 & 9 &0.6& 0.59&0.09&0.09&91.0\\
    9 & 10 &0.59&0.5&0.09&0.09&54.7\\
    10 & 11 &0.5&0.47&0.09&0.11&99.1\\
    11 & 12 &0.47&0.42&0.11&0.11&68.3\\
    12 & 13 &0.42&0.39&0.11&0.04&95.6\\
    13 & 14 &0.39&0.37&0.04&0.09&83.8\\
    14 & 15 &0.37&0.35&0.09&0.10&86.2\\
    15 & 16 &0.35&0.33&0.10&0.10&97.2\\
    16 & 17 &0.33&0.30&0.10&0.05&76.9\\
    17 & 18 &0.30&0.28&0.05&0.14&94.9\\
    18 & 19 &0.28&0.26&0.14&0.16&99.1\\
    19 & 21 (SA) &0.26&0&0.16&0.06&61.0\\
    19 & 22 (SST) &0.26&0&0.16&0.12&79.1\\
  
\end{tabular}
\end{table}

\section{Discussion\label{secDiscussion}}
\subsection{Role of shock waves and compressibility in sustaining buffet}
Summarizing the results of this study that highlight the connection between transonic buffet and incompressible LFO for the NACA 0012 airfoil at $Re = 10^7$: (a) both phenomena show a distinct peak in the power spectrum of the lift coefficient which occurs at a low frequency with $St_b$ in range of $0.05$ to 0.15 (i.e., well below that associated with vortex shedding, Kelvin-Helmholtz instabilities or turbulence \cite{Zauner2020FTaC}), (b) both involve fore-aft chordwise oscillations of the separation or reattachment point, (c) SPOD shows that the pressure fluctuation on the suction side is out of phase with the pressure fluctuation near the trailing edge for both and (d) there is a strong correlation between SPOD modes as flow parameters are gradually varied from transonic to incompressible values. Thus, both qualitative and quantitative relations have been established in the present study to link transonic buffet and incompressible LFO. This connection implies that neither shock waves nor compressibility is necessary to sustain these oscillations, although they might affect the oscillatory features, such as amplitude and onset conditions. We now examine the results from other studies to generalize these conclusions.

\subsection{Consistency with other studies}
The connection between transonic buffet and LFO has also been shown at relatively low Reynolds numbers ($Re \sim O(10^4)$) by Moise \textit{et al.} \cite{Moise2024} for the same NACA 0012 airfoil using LES. The major difference at such low Reynolds numbers is that multiple weak shock waves develop in the flow field, in contrast to the single shock wave observed here. The result that the two phenomena are linked at different Reynolds numbers, irrespective of whether there is a single shock wave or more, is consistent with the current conclusion that the shock wave or its structure is not essential to sustain the oscillations. Similarly, the results from sensitivity and resolvent analysis \cite{Sartor2015, Paladini2019a, Kojima2020} indicate that the buffet flow field is receptive to forcing at the shock foot and that the wave maker of transonic buffet is in the region containing the shock foot. These are consistent with the present study, as the region associated with the shock foot also contains the point of flow separation when strong shock waves develop. Similar conclusions in the other recent studies at low Reynolds number two-dimensional simulations \cite{Fujino2024,DAfiero2025} are also consistent with the present results. Further, the conclusion that incompressible LFO and transonic buffet are linked is also corroborated by the fact that both phenomena arise due to linearly unstable global modes \cite{Crouch2009, Iorio2016}.

\subsection{Implications on feedback loop models}
\label{subSecFeedback}
Although transonic buffet has been clearly shown to arise due to a global instability, the driving and restoring forces that sustain this instability remain unclear. Alternatively, several feedback loop models have been proposed to explain what can drive this instability. The result that shock waves and compressibility are not essential to transonic buffet has important implications for these models. As noted in Sec.~\ref{secIntro}, the popular model of Lee \cite{Lee1990} suggests that waves generated at the shock foot travel downstream to interact with the trailing edge, leading to acoustic waves that travel upstream and interact with shock waves, leading to self-sustained oscillations. The buffet time-period was then predicted as the sum of the time required for downstream traveling waves to reach the trailing edge and the time required for the upstream traveling acoustic waves to reach the shock wave, \textit{i.e.} $T_b = T_\mathrm{down}+T_\mathrm{up}$. Several variants of this model have also been proposed in other studies  (\textit{e.g.}, \cite{erickson1947, mabey1981, Gibb1988, Jacquin2009}). However, all such models rely on the presence of upstream-traveling acoustic waves that interact with the shock wave to induce oscillations. This mechanism, however, fails to explain the presence of buffet-like oscillations under subsonic conditions, where there are no upstream barriers to acoustic wave propagation, such as a shock wave or a supersonic pocket. Furthermore, it has been suggested that feedback mechanisms need not involve shock waves. Instead, an acoustic feedback loop may exist, wherein waves generated at the separation point near the shock foot travel downstream, interact with the trailing edge, and then traverse upstream back to the separation location \cite{Paladini2019a, Fujino2024}. However, the present incompressible results are not consistent even with such feedback models. Since acoustic waves propagate at infinite speed in an incompressible flow, \textit{i.e.}, $T_\mathrm{up} = 0$, this would imply that the downstream traveling waves generated at the separation location will take one buffet cycle to reach the trailing edge (\textit{i.e.}, $T_b = T_\mathrm{down}$). Although footprints of downstream traveling structures are discernible in the space-time diagrams shown in Fig.~\ref{fig:xtLFO}(a) (as highlighted by the dashed lines), the time taken for these waves to reach the trailing edge is very low as compared to the buffet time period ($T_\mathrm{down} \approx 1$ while $T_b \approx 15$). This suggests that the downstream propagating waves that can be observed in both compressible and incompressible simulations (Fig.~\ref{fig:XT_buffet} and Fig.~\ref{fig:xtLFO}) are correlated with the oscillations but are not necessary for the latter's sustenance. Thus, the present results based on incompressible RANS simulations suggest that even feedback models not reliant on shock waves are insufficient to explain the sustained oscillations observed.

\section{Conclusion}
\label{secConclusion}
This study examines the relation between transonic buffet and incompressible low-frequency oscillations (LFO) by performing both compressible and incompressible URANS simulations of the flow over a NACA 0012 airfoil at a moderately high Reynolds number of $Re = 10^7$. Independent simulations with incremental variations in the angle of attack and simultaneous reductions in freestream Mach numbers are carried out over a wide range of angles and Mach numbers. It is shown that oscillations at a low frequency occur at all angles within a narrow range of freestream Mach numbers, including when the flow is transonic, subsonic, or purely incompressible. By analyzing the simulation results, we provide both qualitative and quantitative evidence showing that these oscillations are linked. These include (i) the low frequency of the oscillations ($St \sim O(10^{-1})$), (ii) the chordwise oscillation of the separation point, (iii) visual similarity of the spatial structure of the SPOD modes, and (iv) high correlation values between SPOD modes. Thus, the present study generalizes previous results at low Reynolds numbers and natural transition conditions ($Re \sim O(10^4)$ where multiple shock waves are present in transonic fields) to high Reynolds numbers and fully turbulent conditions ($Re = 10^7$, where a single shock wave is present in transonic fields). 

One main consequence of the present study is that shock waves and compressibility are not essential to sustain buffet. Note that this does not imply that shock waves or compressibility do not have an effect on the oscillations, but that their effects are only secondary. Thus, models that involve feedback mechanisms that require shock waves to be present are inadequate in explaining these oscillations. Further, the separation characteristics observed in the incompressible simulations suggest that even acoustic feedback mechanisms that do not involve shock waves cannot explain the buffet oscillation cycle. Thus, while the current study corroborates the results that both incompressible LFO and transonic buffet arise due to globally unstable modes, the mechanism that drives the instability remains unclear and requires further scrutiny. Further, since transonic buffet and low-frequency oscillations are shown to be linked, the flow control strategies already developed to mitigate one can potentially be exploited for mitigating the other and extending the flight envelope, which requires further exploration.  

A limitation of this study is that it employs URANS simulations, which are not ideal for situations with large degrees of separation. The sensitivity of the simulations to the choice of the turbulence model at high angles of attack and the intermittent nature of the oscillations when Menter's SST $k-\omega$ model is adopted requires further scrutiny using high-fidelity approaches. Similarly, examining instability features using a global linear stability analysis framework can shed light on how the bifurcation characteristics of these oscillations change as one goes from the transonic to the incompressible regime. 

\section*{Appendix}
\label{Appendix}
\subsection*{Effect of turbulence models at high angles of attack}

\begin{figure}[hbt!]
    \centering
        \includegraphics[trim={0cm 0cm 0cm 0cm},  
 clip,width=0.45\textwidth]{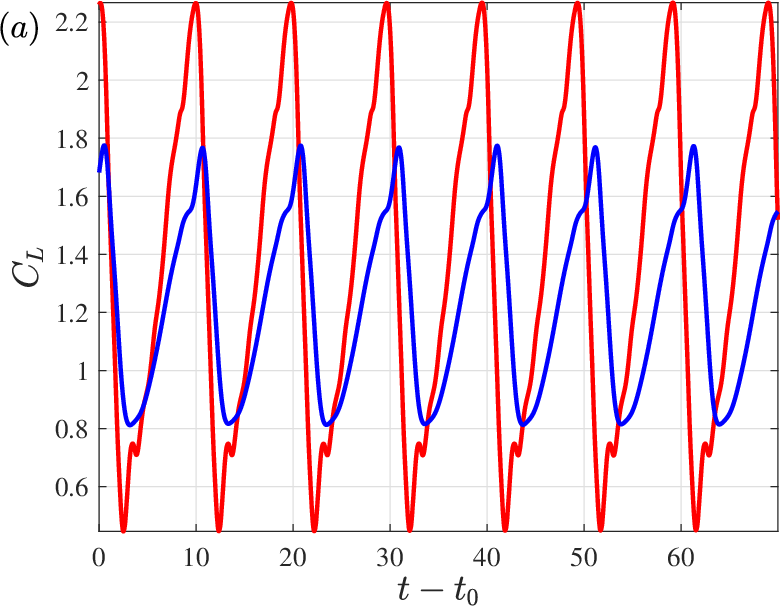}
        \hspace{0.25cm}
        \includegraphics[trim={0cm 0cm 0cm 0cm}, clip,width=0.45\textwidth]{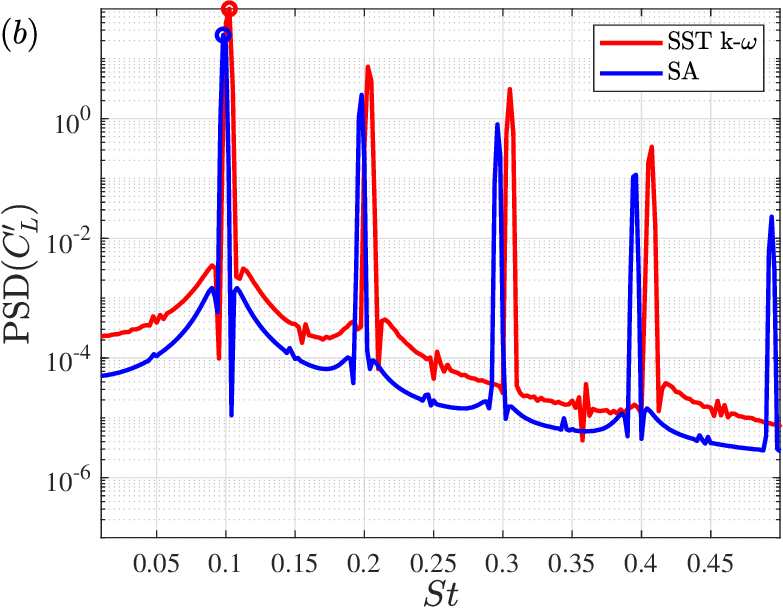}
    \caption{(a) Temporal variation of lift coefficient and (b) power spectra of its fluctuating component in compressible URANS simulations for different turbulence models at $\alpha = 16^\circ$, $M = 0.33$, $Re = 10^7$ for the NACA 0012 airfoil.}
    \label{figApp:SAvSST16}
\end{figure}

As noted in Sec.~\ref{MethodsValid}, the results of buffet are sensitive to the turbulence model used in the URANS simulations at high $\alpha$. This is further investigated in this section. Firstly, we examine the effect of the turbulence model in the compressible simulations. The results for the SA and Menter's SST $k-\omega$ models are compared for the case of $\alpha  = 16^\circ$ and $M = 0.33$ in Fig.~\ref{figApp:SAvSST16}. Under these conditions, the latter turbulence model leads to regular periodic oscillations. The oscillation frequency for the two cases is $St_b \approx 0.1$ with less than $2\%$ difference between the two cases, although there is a significant difference in the oscillation amplitudes. The change in the time-averaged $C_L$ is negligible, with $\bar{C}_L = 1.34$ for the SST $k-\omega$ model and 1.24 for the SA model. Thus, it is seen that the oscillation amplitude is sensitive to the turbulence model but not the oscillation frequency. 

\begin{figure}[t]
    \centering
        \includegraphics[trim={0cm 0cm 0cm 0cm},  
 clip,width=0.47\textwidth]{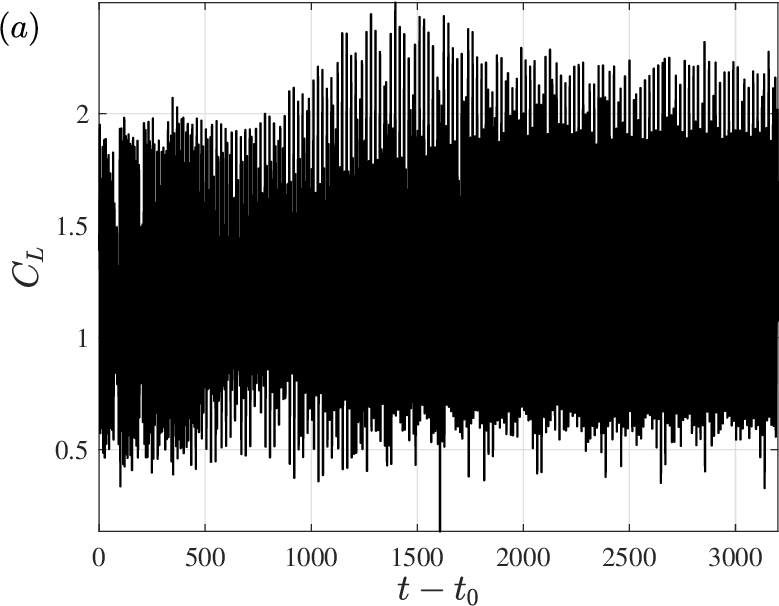}
        \hspace{0.5cm}
        \includegraphics[trim={0cm 0cm 0cm 0cm},  
 clip,width=0.47\textwidth]{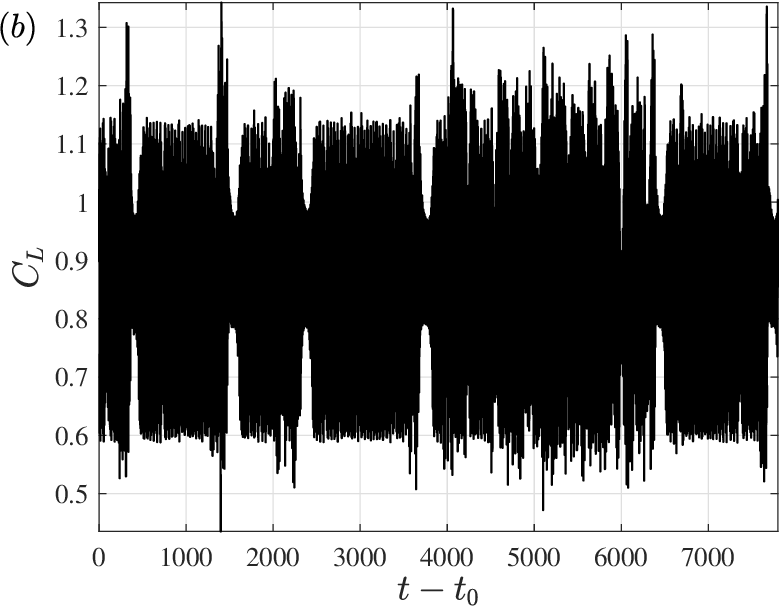}
    \caption{Temporal variation of the lift coefficient past transients obtained from (a) compressible simulations at \(\alpha = 18^\circ\), \(M=0.35\) and (b) incompressible simulations at \(\alpha=22^\circ\) for SST \(k-\omega\) model for NACA 0012 at \(Re = 10^7\).}
    \label{figApp:LongTime}
\end{figure}

For higher angles of attack above $16^\circ$, intermittent long-time variations are observed in the temporal dynamics of the lift coefficient when the SST $k-\omega$ turbulence model is used. This is highlighted for the compressible simulations in Fig.~\ref{figApp:LongTime}(a) for the case of $\alpha = 19^\circ$ and $M = 0.35$. Here, the long-time dynamics show the intermittent nature of the maximum and minimum $C_L$ values attained in the simulations. A similar behavior is observed in the incompressible simulations for the SST $k-\omega$ model, as shown in Fig.~\ref{figApp:LongTime}(b) for $\alpha = 22^\circ$. By contrast, the oscillations are regular for all times for the SA model (not shown).

\section*{Funding Sources}
The computational resources required for analyzing the simulation data were obtained through the Start-up Research Grant (Project number: SRG/2023/000438) from the Science and Engineering Research Board  (currently, Anusandhan National Research Foundation), India. 

\section*{Acknowledgments}
We are grateful to Dr. Markus Zauner and Dr. Fabien Gand for insightful discussions. The support and the resources provided by PARAM Sanganak under the National
Supercomputing Mission, Government of India, and the High-Performance Computing (HPC) cluster at the Indian Institute of Technology, Kanpur, are gratefully acknowledged. 

\bibliography{sample}

\begin{thebibliography}{61}
\newcommand{\enquote}[1]{``#1''}
\providecommand{\natexlab}[1]{#1}
\providecommand{\url}[1]{\texttt{#1}}
\providecommand{\urlprefix}{URL }
\expandafter\ifx\csname urlstyle\endcsname\relax
  \providecommand{\doi}[1]{\discretionary{}{}{}https://doi.org/#1}\else
  \providecommand{\doi}[1]{\discretionary{}{}{}\urlstyle{rm}\url{https://doi.org/#1}}\fi

\bibitem[{Hilton and Fowler(1947)}]{Hilton1947}
Hilton, F., and Fowler, R., \enquote{{Photographs of Shock Wave Movement},} \emph{Natl. Phys. Lab.}, Vol. R\&M 2692, 1947.

\bibitem[{Zaman et~al.(1989)Zaman, McKinzie, and Rumsey}]{Zaman1989}
Zaman, K.~B., McKinzie, D.~J., and Rumsey, C.~L., \enquote{{A Natural Low-Frequency Oscillation of the Flow over an Airfoil Near Stalling Conditions},} \emph{J. Fluid Mech.}, Vol. 202, No. 403, 1989, pp. 403--442.
\newblock \doi{10.1017/S0022112089001230}.

\bibitem[{Dandois et~al.(2018)Dandois, Mary, and Brion}]{Dandois2018}
Dandois, J., Mary, I., and Brion, V., \enquote{{Large-eddy simulation of laminar transonic buffet},} \emph{J. Fluid Mech.}, Vol. 850, 2018, pp. 156--178.
\newblock \doi{10.1017/jfm.2018.470}.

\bibitem[{Zauner and Sandham(2020)}]{Zauner2020FTaC}
Zauner, M., and Sandham, N.~D., \enquote{{Modal Analysis of a Laminar-Flow Airfoil under Buffet Conditions at Re = 500,000},} \emph{Flow, Turbul. Combust.}, Vol. 104, No. 2-3, 2020, pp. 509--532.
\newblock \doi{10.1007/s10494-019-00087-z}.

\bibitem[{Sandham(2008)}]{Sandham2008}
Sandham, N.~D., \enquote{{Transitional separation bubbles and unsteady aspects of aerofoil stall},} \emph{Aeronaut. J.}, Vol. 112, No. 1133, 2008, pp. 395--404.
\newblock \doi{10.1017/S0001924000002359}.

\bibitem[{Almutairi et~al.(2010)Almutairi, Jones, and Sandham}]{Almutairi2010}
Almutairi, J.~H., Jones, L.~E., and Sandham, N.~D., \enquote{{Intermittent bursting of a laminar separation bubble on an airfoil},} \emph{AIAA J.}, Vol.~48, No.~2, 2010, pp. 414--426.
\newblock \doi{10.2514/1.44298}.

\bibitem[{Mons et~al.(2024)Mons, Vervynck, and Marquet}]{Mons2024}
Mons, V., Vervynck, A., and Marquet, O., \enquote{Data assimilation and linear analysis with turbulence modelling: application to airfoil stall flows with PIV measurements,} \emph{Theor. Comput. Fluid Dyn.}, Vol.~38, 2024, pp. 403--429.
\newblock \doi{10.1007/s00162-024-00703-3}.

\bibitem[{McDevitt and Okuno(1985)}]{McDevitt1985}
McDevitt, J.~B., and Okuno, A.~F., \enquote{{Static and Dynamic Pressure Measurements on a NACA 0012 Airfoil in the Ames High Reynolds Number Facility.}} \emph{NASA Tech. Pap.}, 1985.

\bibitem[{Lee(1989)}]{Lee1989}
Lee, B.~H., \enquote{{Investigation of flow separation on a supercritical airfoil},} \emph{J. Aircr.}, Vol.~26, No.~11, 1989, pp. 1032--1037.
\newblock \doi{10.2514/3.45876}.

\bibitem[{Zauner et~al.(2019)Zauner, {De Tullio}, and Sandham}]{Zauner2019}
Zauner, M., {De Tullio}, N., and Sandham, N.~D., \enquote{{Direct numerical simulations of transonic flow around an airfoil at moderate reynolds numbers},} \emph{AIAA J.}, Vol.~57, No.~2, 2019, pp. 597--607.
\newblock \doi{10.2514/1.J057335}.

\bibitem[{Moise et~al.(2024)Moise, Zauner, and Sandham}]{Moise2024}
Moise, P., Zauner, M., and Sandham, N.~D., \enquote{Connecting transonic buffet with incompressible low-frequency oscillations on aerofoils,} \emph{J. Fluid Mech.}, Vol. 981, 2024, p. A23.
\newblock \doi{10.1017/jfm.2023.1065}.

\bibitem[{Hristov and Ansell(2018)}]{Hristov2018}
Hristov, G., and Ansell, P.~J., \enquote{Poststall Hysteresis and Flowfield Unsteadiness on a NACA 0012 Airfoil,} \emph{AIAA J.}, Vol.~56, No.~7, 2018, pp. 2528--2539.
\newblock \doi{10.2514/1.J056774}.

\bibitem[{Kharsansk~Atallah et~al.(2024)Kharsansk~Atallah, Pastur, Monchaux, and Zimmer}]{Atallah2024}
Kharsansk~Atallah, I., Pastur, L., Monchaux, R., and Zimmer, L., \enquote{From low-frequency oscillations to Markovian bistable stall dynamics,} \emph{Phys. Rev. Fluids}, Vol.~9, 2024, p. 063902.
\newblock \doi{10.1103/PhysRevFluids.9.063902}.

\bibitem[{Bernardini et~al.(2016)Bernardini, Benton, Hipp, and Bons}]{Bernardini2016}
Bernardini, C., Benton, S.~I., Hipp, K.~D., and Bons, J.~P., \enquote{Large Low-Frequency Oscillations Initiated by Flow Control on a Poststall Airfoil,} \emph{AIAA J.}, Vol.~54, No.~5, 2016, pp. 1616--1627.
\newblock \doi{10.2514/1.J054321}.

\bibitem[{Iorio et~al.(2016)Iorio, Gonz\'{a}lez, and Mart\'{\i}nez-Cava}]{Iorio2016}
Iorio, M.~C., Gonz\'{a}lez, L.~M., and Mart\'{\i}nez-Cava, A., \enquote{Global Stability Analysis of a Compressible Turbulent Flow Around a High-Lift Configuration,} \emph{AIAA J.}, Vol.~54, No.~2, 2016, pp. 373--385.
\newblock \doi{10.2514/1.J054211}.

\bibitem[{Lee(1990)}]{Lee1990}
Lee, B. H.~K., \enquote{{Oscillatory shock motion caused by transonic shock boundary-layer interaction},} \emph{AIAA J.}, Vol.~28, No.~5, 1990, pp. 942--944.
\newblock \doi{10.2514/3.25144}.

\bibitem[{Iovnovich and Raveh(2012)}]{Iovnovich2012}
Iovnovich, M., and Raveh, D.~E., \enquote{{Reynolds-averaged Navier-Stokes study of the shock-buffet instability mechanism},} \emph{AIAA J.}, Vol.~50, No.~4, 2012, pp. 880--890.
\newblock \doi{10.2514/1.J051329}.

\bibitem[{Giannelis et~al.(2017)Giannelis, Vio, and Levinski}]{Giannelis2017}
Giannelis, N.~F., Vio, G.~A., and Levinski, O., \enquote{{A review of recent developments in the understanding of transonic shock buffet},} \emph{Prog. Aerosp. Sci.}, Vol.~92, 2017, pp. 39--84.
\newblock \doi{10.1016/j.paerosci.2017.05.004}.

\bibitem[{Moise et~al.(2022{\natexlab{a}})Moise, Zauner, and Sandham}]{Moise2022}
Moise, P., Zauner, M., and Sandham, N.~D., \enquote{Large-eddy simulations and modal reconstruction of laminar transonic buffet,} \emph{J. Fluid Mech.}, Vol. 944, 2022{\natexlab{a}}, p. A16.
\newblock \doi{10.1017/jfm.2022.471}.

\bibitem[{Lusher et~al.(2024)Lusher, Sansica, and Hashimoto}]{Lusher2024}
Lusher, D.~J., Sansica, A., and Hashimoto, A., \enquote{Effect of Tripping and Domain Width on Transonic Buffet on Periodic NASA-CRM Airfoils,} \emph{AIAA J.}, Vol.~62, No.~11, 2024, pp. 4411--4430.
\newblock \doi{10.2514/1.J063979}.

\bibitem[{Moise et~al.(2022{\natexlab{b}})Moise, Zauner, Sandham, Timme, and He}]{Moise2022Trip}
Moise, P., Zauner, M., Sandham, N.~D., Timme, S., and He, W., \enquote{Transonic Buffet Characteristics Under Conditions of Free and Forced Transition,} \emph{AIAA J.}, Vol.~0, No.~0, 2022{\natexlab{b}}, pp. 1--16.
\newblock \doi{10.2514/1.J062362}.

\bibitem[{Brion et~al.(2020)Brion, Dandois, Mayer, Reijasse, Lutz, and Jacquin}]{Brion2020}
Brion, V., Dandois, J., Mayer, R., Reijasse, P., Lutz, T., and Jacquin, L., \enquote{{Laminar buffet and flow control},} \emph{Proc. Inst. Mech. Eng. Part G J. Aerosp. Eng.}, Vol. 234, No.~1, 2020, pp. 124--139.
\newblock \doi{10.1177/0954410018824516}.

\bibitem[{Zauner et~al.(2023)Zauner, Moise, and Sandham}]{Zauner2023}
Zauner, M., Moise, P., and Sandham, N., \enquote{On the Co-existence of Transonic Buffet and Separation-Bubble Modes for the OALT25 Laminar-Flow Wing Section,} \emph{Flow Turbul. Combust.}, Vol. 110, No. 1023--1057, 2023.
\newblock \doi{10.1007/s10494-023-00415-4}.

\bibitem[{Zauner et~al.(2024)Zauner, Lusher, Moise, Sansica, Hashimoto, and Sandham}]{Zauner2024}
Zauner, M., Lusher, D.~J., Moise, P., Sansica, A., Hashimoto, A., and Sandham, N.~D., \emph{Open-Source Parametric Airfoils to Study Geometric Effects on Buffet}, AIAA, 2024, pp. 1--34.
\newblock \doi{10.2514/6.2024-3508}.

\bibitem[{Crouch et~al.(2019)Crouch, Garbaruk, and Strelets}]{Crouch2019}
Crouch, J.~D., Garbaruk, A., and Strelets, M., \enquote{{Global instability in the onset of transonic-wing buffet},} \emph{J. Fluid Mech.}, Vol. 881, 2019, pp. 3--22.
\newblock \doi{10.1017/jfm.2019.748}.

\bibitem[{Paladini et~al.(2019{\natexlab{a}})Paladini, Marquet, Sipp, Robinet, and Dandois}]{Paladini2019a}
Paladini, E., Marquet, O., Sipp, D., Robinet, J.~C., and Dandois, J., \enquote{{Various approaches to determine active regions in an unstable global mode: Application to transonic buffet},} \emph{J. Fluid Mech.}, Vol. 881, No.~M, 2019{\natexlab{a}}, pp. 617--647.
\newblock \doi{10.1017/jfm.2019.761}.

\bibitem[{Dandois(2016)}]{Dandois2016}
Dandois, J., \enquote{{Experimental study of transonic buffet phenomenon on a 3D swept wing},} \emph{Phys. Fluids}, Vol.~28, No.~1, 2016.
\newblock \doi{10.1063/1.4937426}.

\bibitem[{Timme(2020)}]{Timme2020}
Timme, S., \enquote{{Global instability of wing shock-buffet onset},} \emph{J. Fluid Mech.}, Vol. 885, 2020, pp. 1--32.
\newblock \doi{10.1017/jfm.2019.1001}.

\bibitem[{Jacquin et~al.(2009)Jacquin, Molton, Deck, Maury, and Soulevant}]{Jacquin2009}
Jacquin, L., Molton, P., Deck, S., Maury, B., and Soulevant, D., \enquote{{Experimental study of shock oscillation over a transonic supercritical profile},} \emph{AIAA J.}, Vol.~47, No.~9, 2009, pp. 1985--1994.
\newblock \doi{10.2514/1.30190}.

\bibitem[{Lusher et~al.(2025)Lusher, Sansica, and Hashimoto}]{Lusher2025}
Lusher, D.~J., Sansica, A., and Hashimoto, A., \enquote{Implicit large eddy simulations of three-dimensional turbulent transonic buffet on wide-span infinite wings,} \emph{J. Fluid Mech.}, Vol. 1007, 2025, p. A26.
\newblock \doi{10.1017/jfm.2025.48}.

\bibitem[{Busquet et~al.(2021)Busquet, Marquet, Richez, Juniper, and Sipp}]{Busquet2021}
Busquet, D., Marquet, O., Richez, F., Juniper, M., and Sipp, D., \enquote{{Bifurcation scenario for a two-dimensional static airfoil exhibiting trailing edge stall},} \emph{J. Fluid Mech.}, Vol. 928, 2021, pp. 1--27.
\newblock \doi{10.1017/jfm.2021.767}.

\bibitem[{Rinoie and Takemura(2004)}]{rinoie_takemura_2004}
Rinoie, K., and Takemura, N., \enquote{Oscillating behaviour of laminar separation bubble formed on an aerofoil near stall,} \emph{Aeronaut. J.}, Vol. 108, No. 1081, 2004, p. 153–163.
\newblock \doi{10.1017/S0001924000151607}.

\bibitem[{Aniffa and Mandal(2023{\natexlab{a}})}]{Aniffa2023a}
Aniffa, S.~M., and Mandal, A.~C., \enquote{Experiments on the unsteady massive separation over an aerofoil,} \emph{Phys. Rev. Fluids}, Vol.~8, 2023{\natexlab{a}}, p. 123901.
\newblock \doi{10.1103/PhysRevFluids.8.123901}.

\bibitem[{Aniffa and Mandal(2023{\natexlab{b}})}]{Aniffa2023b}
Aniffa, S.~M., and Mandal, A.~C., \enquote{Experiments on the low-frequency oscillation of a separated shear layer,} \emph{Phys. Rev. Fluids}, Vol.~8, 2023{\natexlab{b}}, p. 023902.
\newblock \doi{10.1103/PhysRevFluids.8.023902}.

\bibitem[{Eljack(2024)}]{Eljack_2024}
Eljack, E.~M., \enquote{The structure and dynamics of the laminar separation bubble,} \emph{J. Fluid Mech.}, Vol. 998, 2024, p. A56.
\newblock \doi{10.1017/jfm.2024.855}.

\bibitem[{Paladini et~al.(2019{\natexlab{b}})Paladini, Beneddine, Dandois, Sipp, and Robinet}]{Paladini2019}
Paladini, E., Beneddine, S., Dandois, J., Sipp, D., and Robinet, J.~C., \enquote{{Transonic buffet instability: From two-dimensional airfoils to three-dimensional swept wings},} \emph{Phys. Rev. Fluids}, Vol.~4, No.~10, 2019{\natexlab{b}}, p. 103906.
\newblock \doi{10.1103/PhysRevFluids.4.103906}.

\bibitem[{Deck(2005)}]{Deck2005}
Deck, S., \enquote{{Numerical simulation of transonic buffet over a supercritical airfoil},} \emph{AIAA J.}, Vol.~43, No.~7, 2005, pp. 1556--1566.
\newblock \doi{10.2514/1.9885}.

\bibitem[{Xiao et~al.(2006)Xiao, Tsai, and Liu}]{Xiao2006a}
Xiao, Q., Tsai, H.~M., and Liu, F., \enquote{{Numerical study of transonic buffet on a supercritical airfoil},} \emph{AIAA J.}, Vol.~44, No.~3, 2006, pp. 620--628.
\newblock \doi{10.2514/1.16658}.

\bibitem[{Ranjan~Majhi and Venkatraman(2023)}]{Mahji2023}
Ranjan~Majhi, J., and Venkatraman, K., \enquote{On the Nature of Transonic Shock Buffet in an Axial-Flow Fan,} \emph{AIAA J.}, Vol.~61, No.~12, 2023, pp. 5390--5403.
\newblock \doi{10.2514/1.J063318}.

\bibitem[{Crouch et~al.(2007)Crouch, Garbaruk, and Magidov}]{Crouch2007}
Crouch, J.~D., Garbaruk, A., and Magidov, D., \enquote{{Predicting the onset of flow unsteadiness based on global instability},} \emph{J. Comput. Phys.}, Vol. 224, No.~2, 2007, pp. 924--940.
\newblock \doi{10.1016/j.jcp.2006.10.035}.

\bibitem[{Crouch et~al.(2009)Crouch, Garbaruk, Magidov, and Travin}]{Crouch2009}
Crouch, J.~D., Garbaruk, A., Magidov, D., and Travin, A., \enquote{{Origin of transonic buffet on aerofoils},} \emph{J. Fluid Mech.}, Vol. 628, 2009, pp. 357--369.
\newblock \doi{10.1017/S0022112009006673}.

\bibitem[{Crouch et~al.(2024)Crouch, Ahrabi, and Kamenetskiy}]{Crouch2024}
Crouch, J., Ahrabi, B., and Kamenetskiy, D., \enquote{Weakly nonlinear behaviour of transonic buffet on airfoils,} \emph{J. Fluid Mech.}, Vol. 999, 2024, p.~A8.
\newblock \doi{10.1017/jfm.2024.499}.

\bibitem[{Menter(1994)}]{Menter}
Menter, F.~R., \enquote{Two-equation eddy-viscosity turbulence models for engineering applications,} \emph{AIAA J.}, Vol.~32, No.~8, 1994, pp. 1598--1605.
\newblock \doi{10.2514/3.12149}.

\bibitem[{Spalart(2000)}]{Spalart}
Spalart, P., \emph{Trends in turbulence treatments}, AIAA, 2000, pp. 2000--2306.
\newblock \doi{10.2514/6.2000-2306}.

\bibitem[{Giannelis et~al.(2018)Giannelis, Levinski, and Vio}]{Giannelis2018}
Giannelis, N.~F., Levinski, O., and Vio, G.~A., \enquote{{Influence of Mach number and angle of attack on the two-dimensional transonic buffet phenomenon},} \emph{Aerosp. Sci. Technol.}, Vol.~78, 2018, pp. 89--101.
\newblock \doi{10.1016/j.ast.2018.03.045}.

\bibitem[{Roe(1986)}]{Roe}
Roe, P.~L., \enquote{Characteristic-Based Schemes for the Euler Equations,} \emph{Annu. Rev. Fluid Mech.}, Vol.~18, 1986, pp. 337--365.
\newblock \doi{https://doi.org/10.1146/annurev.fl.18.010186.002005}.

\bibitem[{Lumley(1970)}]{Lumley1970}
Lumley, J.~L., \emph{Stochastic tools in turbulence}, 1\textsuperscript{st} ed., Academic Press, 1970.

\bibitem[{Glauser et~al.(1987)Glauser, Leib, and George}]{Glauser1987}
Glauser, M.~N., Leib, S.~J., and George, W.~K., \enquote{Coherent Structures in the Axisymmetric Turbulent Jet Mixing Layer,} \emph{Turbulent Shear Flows 5}, edited by F.~Durst, B.~E. Launder, J.~L. Lumley, F.~W. Schmidt, and J.~H. Whitelaw, Springer Berlin Heidelberg, Berlin, Heidelberg, 1987, pp. 134--145.

\bibitem[{Towne et~al.(2018)Towne, Schmidt, and Colonius}]{Towne2018}
Towne, A., Schmidt, O.~T., and Colonius, T., \enquote{{Spectral proper orthogonal decomposition and its relationship to dynamic mode decomposition and resolvent analysis},} \emph{J. Fluid Mech.}, Vol. 847, 2018, pp. 821--867.
\newblock \doi{10.1017/jfm.2018.283}.

\bibitem[{Schmidt and Towne(2019)}]{SCHMIDT201998}
Schmidt, O.~T., and Towne, A., \enquote{An efficient streaming algorithm for spectral proper orthogonal decomposition,} \emph{Comput. Phys. Comm.}, Vol. 237, 2019, pp. 98--109.
\newblock \doi{10.1016/j.cpc.2018.11.009}.

\bibitem[{McDevitt et~al.(1976)McDevitt, Levy, and Deiwert}]{McDevitt1976}
McDevitt, J.~B., Levy, L.~L., and Deiwert, G.~S., \enquote{{Transonic flow about a thick circular-arc airfoil},} \emph{AIAA J.}, Vol.~14, No.~5, 1976, pp. 606--613.
\newblock \doi{10.2514/3.61402}.

\bibitem[{Lee(2001)}]{Lee2001}
Lee, B. H.~K., \enquote{{Self-sustained shock oscillations on airfoils at transonic speeds},} \emph{Prog. Aerosp. Sci.}, Vol.~37, No.~2, 2001, pp. 147--196.
\newblock \doi{10.1016/S0376-0421(01)00003-3}.

\bibitem[{B.McCullough and Gault(1951)}]{McCullough1951}
B.McCullough, G., and Gault, D.~E., \enquote{{Examples of Three Representative Types of Airfoil-section Stall at Low Speed},} Tech. Rep. 2502, NACA TN, 1951.
\newblock \urlprefix\url{https://ntrs.nasa.gov/citations/19930083422}.

\bibitem[{Sartor et~al.(2015)Sartor, Mettot, and Sipp}]{Sartor2015}
Sartor, F., Mettot, C., and Sipp, D., \enquote{{Stability, receptivity, and sensitivity analyses of buffeting transonic flow over a profile},} \emph{AIAA J.}, Vol.~53, No.~7, 2015, pp. 1980--1993.
\newblock \doi{10.2514/1.J053588}.

\bibitem[{Grossi et~al.(2014)Grossi, Braza, and Hoarau}]{Grossi2014}
Grossi, F., Braza, M., and Hoarau, Y., \enquote{{Prediction of transonic buffet by delayed detached-eddy simulation},} \emph{AIAA J.}, Vol.~52, No.~10, 2014, pp. 2300--2312.
\newblock \doi{10.2514/1.J052873}.

\bibitem[{Kojima et~al.(2020)Kojima, Yeh, Taira, and Kameda}]{Kojima2020}
Kojima, Y., Yeh, C.-A., Taira, K., and Kameda, M., \enquote{Resolvent analysis on the origin of two-dimensional transonic buffet,} \emph{J. Fluid Mech.}, Vol. 885, 2020, p.~R1.
\newblock \doi{10.1017/jfm.2019.992}.

\bibitem[{Fujino and Suzuki(2024)}]{Fujino2024}
Fujino, K., and Suzuki, K., \enquote{Mechanism of periodic oscillation in low-Reynolds-number buffet around an airfoil at angle of attack 0,} \emph{Physics of Fluids}, Vol.~36, No.~4, 2024, p. 046126.
\newblock \doi{10.1063/5.0201260}.

\bibitem[{D'Afiero(2025)}]{DAfiero2025}
D'Afiero, F.~M., \enquote{Feature based analysis of low frequency oscillations in transonic low Reynolds airfoil flows,} \emph{Physics of Fluids}, Vol.~37, No.~5, 2025, p. 056112.
\newblock \doi{10.1063/5.0266901}.

\bibitem[{Erickson and Stephenson(1947)}]{erickson1947}
Erickson, A.~L., and Stephenson, J.~D., \enquote{A suggested method of analyzing for transonic flutter of control surfaces based on available experimental evidence,} Tech. Rep. NACA-RM-A7F30, NACA, 1947.
\newblock \urlprefix\url{https://ntrs.nasa.gov/citations/19930085716}.

\bibitem[{Mabey et~al.(1981)Mabey, Welsh, and Cripps}]{mabey1981}
Mabey, D., Welsh, B., and Cripps, B., \emph{Periodic flows on a rigid 14\% thick biconvex wing at transonic speeds}, Royal Aircraft Establishment, 1981.

\bibitem[{Gibb(1988)}]{Gibb1988}
Gibb, J., \enquote{{The Cause and Cure of Periodic Flows at Transonic Speeds},} \emph{Proc. ICAS}, 1988, pp. 1522--1530.

\end{thebibliography}

\end{document}